\documentclass[acmsmall]{acmart}

\AtBeginDocument{%
  }

\setcopyright{acmlicensed}
\copyrightyear{2026}
\acmYear{2026}
\acmDOI{10.1145/3833426}

\acmJournal{TACO}
\acmVolume{23}
\acmNumber{3}
\acmArticle{114}
\acmMonth{9}

\usepackage{multirow}

\usepackage{tikz}
\usepackage[hypcap=false]{caption}
\usepackage{tcolorbox}
\usepackage{soul}
\tcbuselibrary{breakable,skins}

\newlength\szg     \newcommand \hquan[1]{%
\settoheight\szg{#1}%
\tikz[baseline]{
\pgfmathparse{1}
\let\hfs\pgfmathresult
\filldraw[gray] (0,\szg/2) circle (\szg/2+0.35ex);
\node[white] at (0,\szg/2) {\makebox[0em][c]{\scalebox{\hfs}[1]{\textbf{#1}}}};
}}

\newcommand{\revisenew}[1]{#1}
\soulregister\cite7
\soulregister\ref7
\soulregister\autoref7
\soulregister\sectionautorefname7
\soulregister\figureautorefname7
\soulregister\tableautorefname7

\newcommand{\archname}{GUMPIM}

\renewcommand{\sectionautorefname}{Section}
\renewcommand{\figureautorefname}{Fig.}

\begin{document}

\title{\archname{}: Unitary and Malleable Memory for Processing-in-Memory with Guaranteed PIM Pages}
\titlenote{Extension of Conference Paper \cite{um-pim10609641}, this submission adds 30\% new material over the prior conference paper. This work was partially supported by the National Key Research and
Development Program of China (No. 2024YFE0204300), the National Program for Support of Top-notch Young Professionals of China (No. 24Z130300727), National Natural Science Foundation of China (No. 62402311, 62072262, 61834006), Natural Science Foundation
of Shanghai (No. 24ZR1433700), and Key Research and Development Program of Shanghai (No. 25LN3201200).}



\author{Yilong~Zhao}
\email{sjtuzyl@sjtu.edu.cn}
\orcid{0000-0001-8291-6896}
\affiliation{%
  \institution{Shanghai Jiao Tong University}
  \city{Shanghai}
  \state{Shanghai}
  \country{China}
}
\affiliation{%
  \institution{Shanghai Qi Zhi Institute}
  \city{Shanghai}
  \state{Shanghai}
  \country{China}
}

\author{Fangxin~Liu}
\authornote{For correspondence, please contact Li Jiang and Fangxin Liu.}
\orcid{0000-0002-8769-293X}
\email{liufangxin@sjtu.edu.cn}
\affiliation{%
  \institution{Shanghai Jiao Tong University}
  \city{Shanghai}
  \state{Shanghai}
  \country{China}
}
\affiliation{%
  \institution{Shanghai Qi Zhi Institute}
  \city{Shanghai}
  \state{Shanghai}
  \country{China}
}

\author{Yiwei~Hu}
\orcid{0009-0000-1535-8046}
\email{arikara666@sjtu.edu.cn}
\affiliation{%
  \institution{Shanghai Jiao Tong University}
  \city{Shanghai}
  \state{Shanghai}
  \country{China}
}
\affiliation{%
  \institution{Shanghai Qi Zhi Institute}
  \city{Shanghai}
  \state{Shanghai}
  \country{China}
}

\author{Zongwu~Wang}
\orcid{0009-0003-2157-4927}
\email{wangzongwu@sjtu.edu.cn}
\affiliation{%
  \institution{Shanghai Jiao Tong University}
  \city{Shanghai}
  \state{Shanghai}
  \country{China}
}
\affiliation{%
  \institution{Shanghai Qi Zhi Institute}
  \city{Shanghai}
  \state{Shanghai}
  \country{China}
}

\author{Mingyu~Gao}
\orcid{0000-0001-8433-7281}
\email{gaomy@tsinghua.edu.cn}
\affiliation{%
  \institution{Tsinghua University}
  \city{Beijing}
  \state{Beijing}
  \country{China}
}
\affiliation{%
  \institution{Shanghai Qi Zhi Institute}
  \city{Shanghai}
  \state{Shanghai}
  \country{China}
}

\author{Onur~Mutlu}
\orcid{0000-0002-0075-2312}
\email{omutlu@gmail.com}
\affiliation{%
  \institution{ETH Zurich}
  \city{Zurich}
  \state{Zurich}
  \country{Switzerland}
}

\author{He Xian}
\orcid{0009-0008-2099-4109}
\email{51265900021@stu.ecnu.edu.cn}
\affiliation{%
  \institution{Shanghai Qi Zhi Institute}
  \city{Shanghai}
  \state{Shanghai}
  \country{China}
}

\author{Hanlin Dong}
\orcid{0000-0002-0075-2312}
\email{51265900020@stu.ecnu.edu.cn}
\affiliation{%
  \institution{Shanghai Qi Zhi Institute}
  \city{Shanghai}
  \state{Shanghai}
  \country{China}
}

\author{Naifeng Jing}
\orcid{0000-0001-8588-9483}
\email{sjtuj@sjtu.edu.cn}
\affiliation{%
  \institution{Shanghai Jiao Tong University}
  \city{Shanghai}
  \state{Shanghai}
  \country{China}
}

\author{Xiaoyao Liang}
\orcid{0000-0002-2790-5884}
\email{liang-xy@cs.sjtu.edu.cn} 
\affiliation{%
  \institution{Shanghai Jiao Tong University}
  \city{Shanghai}
  \state{Shanghai}
  \country{China}
}

\author{Haibing Guan}
\orcid{0000-0002-4714-7400}
\email{hbguan@sjtu.edu.cn} 
\affiliation{%
  \institution{Shanghai Jiao Tong University}
  \city{Shanghai}
  \state{Shanghai}
  \country{China}
}

\author{Qiang Xu}
\orcid{0000-0001-6747-126X}
\email{xuqiang@cuhk.edu.hk}
\affiliation{%
  \institution{The Chinese University of Hong Kong}
  \city{Hong Kong}
  \state{Hong Kong}
  \country{China}
}

\author{Chixiao Chen}
\orcid{0000-0002-5980-4236}
\email{cxchen@fudan.edu.cn}
\affiliation{%
  \institution{Fudan University}
  \city{Shanghai}
  \state{Shanghai}
  \country{China}
}

\author{Shuxue Quan}
\orcid{}
\email{pine.quan@yizhu-tech.com}
\affiliation{%
  \institution{Yizhu Technology Co., Ltd.}
  \city{Suzhou}
  \state{Jiangsu}
  \country{China}
}

\author{Tao~Yang}
\orcid{0000-0001-8588-9483}
\email{yt594584152@sjtu.edu.cn}
\affiliation{%
  \institution{Huawei Technologies Co., Ltd.}
  \city{Shenzhen}
  \state{Guangdong}
  \country{China}
}
\affiliation{%
  \institution{Shanghai Jiao Tong University}
  \city{Shanghai}
  \state{Shanghai}
  \country{China}
}
\affiliation{%
  \institution{Shanghai Qi Zhi Institute}
  \city{Shanghai}
  \state{Shanghai}
  \country{China}
}
\author{Li Jiang}
\authornotemark[2]
\orcid{0000-0002-7353-8798}
\email{ljiang_cs@sjtu.edu.cn}
\affiliation{%
  \institution{Shanghai Jiao Tong University}
  \city{Shanghai}
  \state{Shanghai}
  \country{China}
}
\affiliation{%
  \institution{Shanghai Qi Zhi Institute}
  \city{Shanghai}
  \state{Shanghai}
  \country{China}
}

\renewcommand{\shortauthors}{Zhao et al.}

\begin{abstract}
  DRAM-based Processing-in-Memory (PIM) addresses the ``memory wall'' by executing computations directly inside main memory.
  However, memory interleaving and virtual memory limit contiguous data size visible to PIM units, constraining PIM task granularity.
  Fine-grained PIM tasks incur significant offloading overhead that negates PIM performance benefits.
  To mitigate this, existing PIM systems drastically isolate PIM memory or disable memory interleaving.
  These design choices, however, decrease the CPU memory bandwidth and introduce extra data transfer, leading to an additional ``system memory wall'' that degrades CPU performance and must be resolved to realize PIM's full potential.

  In this work, we propose \archname{}, a \underline{PIM} system that allows interleaved CPU pages and non-interleaved PIM pages to coexist in a \underline{U}nitary and \underline{M}alleable memory space with \underline{G}uaranteed PIM page allocation.
  \archname{} enables zero-copy during PIM task offloading and maintains CPU memory bandwidth while ensuring low PIM offloading overhead.
  First, we propose a dual-track memory management mechanism consisting of independent page allocation and address translation for CPU and PIM pages.
  Second, we design \archname{} interface hardware on PIM-enabled DRAMs to provide a dynamic address mapping for the different data layouts of CPU and PIM pages.
  Third, we propose a PIM-assisted page migration mechanism that transparently migrates pages while preserving CPU access bandwidth, thereby enabling guaranteed and accelerated PIM page allocation. 
  \archname{} requires no changes to commodity DRAM standards; all hardware modifications are limited to the DRAM side, ensuring full compatibility with existing CPUs and enabling immediate deployment on current HBMx- and LPDDRx-based PIM platforms.
  Our results show only <0.1\% performance degradation for CPU workloads on \archname{}, in contrast to the 25.8\% degradation on PIM systems with memory interleaving turned off.
  For PIM workloads, \archname{} reduces memory allocation and CPU-part computation times by 2.7\texttimes{} and 4.93\texttimes{}, respectively, yielding an end-to-end 2.3\texttimes{} speedup over a state-of-the-art baseline system.
\end{abstract}



\received{25 Mar 2026}
\received[revised]{18 May 2026}
\received[accepted]{21 May 2026}

\maketitle

\section{Introduction}
\label{sec:introduction}

DRAM-based Processing-in-Memory (PIM) is an emerging technology to overcome the ``memory wall'' \cite{wulf95_memory_wall} between processing units and main memory, by integrating computational components (PIM units) into DRAM.
The PIM units are distributed in various levels of DRAM organization including arrays \cite{SIMDRAM10.1145/3445814.3446749,Ambit10.1145/3123939.3124544,mimdram10476478,rowclone_10.1145/2540708.2540725,Functionally-Complete_10476435,salp6237032,proteus_10.1145/3721145.3730420}, banks \cite{UPMEM8875680,FIMDRAM9365862,AiMBench10.1093/bioinformatics/btad155,Aquabolt-XL-9567191,prim9771457,pim-ml10158216,prim2_9651614} or ranks \cite{AxDIMM10.1145/3533737.3535093,AsyncDIMM10946818,CoNDA_8980366}.
These PIM units can directly access data in their own memory device, bypassing the global memory bus and achieving much higher bandwidth than CPUs, making them suitable for memory-intensive operations.
The PIM units are usually customized to accelerate specific memory-bottlenecked applications, such as graph computing, artificial intelligence, databases, genome analysis, and recommendation systems \cite{MetaNMP10.1145/3579371.3589091,PIM-HBM9073325,GDDR6-AiM9895629,TETRIS10.1145/3037697.3037702,tensordimm10.1145/3352460.3358284,RecNMP9138955,Tesseract7284059,google_nn_2021_9563028,sparsep_9912078,google_workloads_10.1145/3173162.3173177,polynesia9835628,grim-filter}.
General-purpose PIM systems emerge to support a broader range of workloads \cite{UPMEM8875680,AxDIMM10.1145/3533737.3535093,pidram10.1145/3563697,ALPHA-PIM-11242082,pim-opt10807322,BIMSA_10.1093/bioinformatics/btae631,pygim_10.1145/3700434,swiftrl_10590059,Homomorphic_10289253, sisa10.1145/3466752.3480133,pim_is_all_you_need10.1145/3676641.3716267,simplepim_10.1109/PACT58117.2023.00017,transpimlib_10158230}.
These systems allow memory-bound program segments to be offloaded to PIM units as \emph{PIM tasks} while keeping compute-bound segments executed by CPUs.
Therefore, the compatibility of these PIM systems with host CPUs becomes critical to ensure cooperation between CPUs and PIM units.

\begin{figure}[h]
    \centering
    \includegraphics[width=0.95\linewidth]{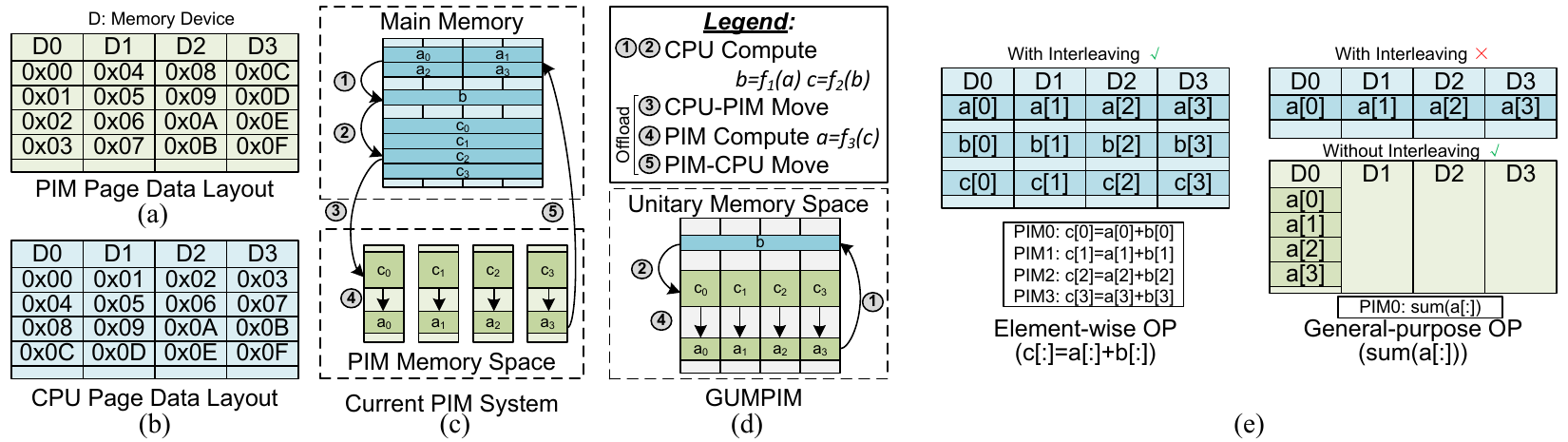}
    \vspace{-10pt}
    \caption{ (a) Data layout of PIM pages and (b) CPU pages.
    (c) Isolated memory space design in existing PIM systems. (d) Proposed unitary \& malleable memory space.
    (e) Element-wise and general-purpose operations with/without memory interleaving.
    }
    \vspace{-15pt}
    \label{fig:isolation}
    \Description{This figure illustrates the isolation and unitary \& malleable memory space design in existing PIM systems and the proposed PIM system.}
\end{figure}

Modern CPU-centric memory systems prevent PIM units from accessing large physically contiguous data due to two mechanisms:
First, in \textbf{virtual memory}, the operating system (OS) maps contiguous virtual addresses to non-contiguous physical pages.
Second, through \textbf{memory interleaving}, the system stripes consecutive cache lines across multiple channels, ranks, or banks (\figureautorefname{}~\ref{fig:isolation}(b)), further limiting data contiguity for individual PIM units.
Together, CPU-centric mechanisms severely limit the effective granularity of PIM tasks.
For example, PIM units in a DIMM DRAM bank cannot even access consecutive bytes within a data structure.

When offloading PIM tasks, the CPU needs to take a series of actions, including context switching \cite{PIMProf9774560} and locking memory regions \cite{UPMEM-Toolchain}.
For certain general-purpose PIM systems, this offloading overhead is even more than 50$\mu$s \cite{UPMEM8875680}.
Such offloading overhead results in PIM units having to process more data at each offload to amortize the cost.
This requires PIM units to have access to even larger contiguous regions of data (like in \figureautorefname{}~\ref{fig:isolation}(a)), further contradicting the design of modern CPU-centric memory systems.

To ensure PIM efficiency, existing systems resort to extreme measures that incur significant overhead or degrade CPU performance.
(i) \textbf{Isolated memory space.}
A dedicated PIM memory space is physically isolated from main memory, as shown in \figureautorefname{} \ref{fig:isolation}(c) \cite{UPMEM8875680,topim10.1145/3470496.3527431,pim_malloc_11408588}.
The CPU accesses this space via physical addresses, enabling manual placement of data on specific memory modules.
However, PIM task offloading requires explicit data transfers between the two spaces over the memory bus (Steps~\hquan{3} and~\hquan{5}), introducing an additional ``memory wall.''
(ii) \textbf{Software data re-layout.}
CPU software explicitly reorganizes data layout to place contiguous blocks within a single memory module \cite{UPMEM8875680,AxDIMM10.1145/3533737.3535093,PIM-HBM9073325,topim10.1145/3470496.3527431,newton_9251855}.
This further increases the offloading overhead -- over 90\% on tasks with frequent, fine-grained data re-layout as in BFS and NW, as evaluated on a real PIM system \cite{prim9771457}.
(iii) \textbf{Globally disabling memory interleaving.}
Disabling memory interleaving can reduce the data re-layout overhead at rank and channel levels \cite{UPMEM8875680,AxDIMM10.1145/3533737.3535093,NDP7429299}.
However, doing so greatly degrades CPU memory bandwidth, aggravating the ``memory wall'' between CPU and DRAM.

To truly break the ``walls'', a unitary, malleable memory space is essential for general-purpose heterogeneous CPU-PIM systems, as depicted in \figureautorefname{}~\ref{fig:isolation}(d).
CPU and PIM pages, with different data layouts (\figureautorefname{} \ref{fig:isolation}(a, b)), coexist in the memory space.
This memory management strategy fulfills the requirements of PIM units for contiguous data layout while avoiding the negative consequences of the above extreme measures.
First, the memory space enables zero-copy by eliminating data transfers (no \hquan{3} and \hquan{5} in \figureautorefname{}~\ref{fig:isolation}(d)) during offloading.
Second, CPUs access PIM pages efficiently and transparently without requiring software-level address remapping and data re-layout.
Third, CPU bandwidth is preserved as in CPU-centric systems because CPU pages can still leverage standard memory interleaving.

To preserve compatibility with existing CPU hardware and OS, the following challenges should be addressed.
(i) \textbf{CPU and PIM page coexistence.}
A unitary memory space requires the PIM unit to share the CPU-managed virtual memory, necessitating that adjacent virtual PIM pages be kept locally in PIM memory.
(ii) \textbf{Different data layouts.} An efficient address mapping is necessary to support CPU and PIM pages with different data layouts.
(iii) \textbf{CPU accessing PIM pages.} Efforts should be made to alleviate the bank conflicts when the host CPU accesses the non-interleaved PIM pages, which potentially degrades CPU memory bandwidth.
(iv) \textbf{Compatibility.} Hardware modifications should primarily focus on the DRAM side to maintain compatibility with existing CPUs and motherboards.

In this work, we propose \archname{}, a general-purpose PIM system with a unitary, malleable memory space and dual-track memory management for CPU and PIM pages.
Existing memory allocation and address mapping mechanisms for CPU pages are kept unchanged to preserve compatibility.
For PIM pages, we propose chunk-based memory management to enable locality for PIM pages and maintain low address translation overhead.
We propose \archname{} interface hardware for efficient address mapping and improve memory bandwidth available to the CPU when it accesses CPU and PIM pages.
Additionally, chunk-based memory management requires allocating contiguous memory for PIM pages, which existing CPU systems lack because memory fragmentation leaves only scattered, non-contiguous free regions \cite{Contiguitas10.1145/3579371.3589079,fragment_numa183961,hawkeye10.1145/3297858.3304064}.
We propose a PIM-assisted page migration mechanism to enable contiguous memory allocation for PIM pages.
All our hardware modifications are limited to the DRAM side.

An earlier version of this work (UM-PIM) was published in the proceedings of ISCA 2024 \cite{um-pim10609641}. 
This journal version significantly extends the previous work with the following new mechanisms and results: 
(1) A PIM-assisted page migration mechanism that enables successful allocation of PIM pages while maintaining full compatibility with existing CPU-side hardware, and significantly accelerates the page allocation (\sectionautorefname{}~\ref{sec:chunk_allocation}).
(2) A generalized design for dual-track memory management extended beyond the original architecture to support diverse DRAM types, including LPDDRx and HBMx, with complete implementation details that were previously only discussed at a high level (\sectionautorefname{}~\ref{sec:other_dram}).
(3) Evaluation of the memory allocation speedup with the proposed PIM-assisted page migration mechanism, and demonstration of the effectiveness of the unitary memory design on LPDDRx- and HBMx-based DRAM PIM (\sectionautorefname{}~\ref{sec:evaluation:chunk_allocation}-\ref{sec:evaluation:other_dram}).
Consistent with the design principle of the original UM-PIM \cite{um-pim10609641}, these new techniques also confine hardware modifications strictly to the DRAM side and ensure compatibility with the CPU-side hardware and DRAM standard protocol. 

The main contributions of this paper are:
\begin{itemize}
    \item We identify the root cause of excessive data orchestration overhead in existing processing-in-memory systems: the additional memory wall imposed by isolated PIM memory spaces that are not directly accessible by the host CPU.
    \item We propose \archname{}, a general-purpose PIM system with a unitary and malleable memory space, which incorporates dual-track memory management along with dedicated hardware support. We demonstrate that this approach naturally extends to other DRAM standards, including HBMx and LPDDRx, without PIM-side modifications.
    \item We design a high-level programming interface with collective communication primitives inspired by NCCL~\cite{NCCL}. By exploiting the shared address space and fine-grained inter-PIM connectivity of \archname{}, these primitives dramatically reduce synchronization and data movement overhead across PIM units.
    \item We introduce a PIM-assisted transparent page migration mechanism that dynamically constructs large contiguous physical memory regions for PIM workloads, eliminating host-side fragmentation and enabling efficient and correct huge-page allocation.
\end{itemize} 

Our comprehensive evaluation results show that the dual-track memory management of \archname{} provides 1.96\texttimes{} speedup on PIM workloads with only <0.1\% performance degradation for CPU workloads, over a state-of-the-art baseline system. The PIM-assisted page migration further reduces the memory allocation overhead by 2.7\texttimes{}.
\section{Background}
\label{sec:background}

\subsection{Address Mapping and Memory Interleaving }
\label{sec:background:interleave}

In modern systems, the virtual address (VAddr) is mapped to the DRAM hardware location in two steps.
First, the memory management unit (MMU) translates the virtual address to the physical address (PAddr).
The virtual page number (VPN) is translated to the physical page number (PPN) by the OS-managed page table, while the page offset is kept unchanged.
After that, the memory controller maps the physical address to the hardware address (HWAddr) using the address mapping rule saved in the BIOS-defined target address decoder (TAD) register, which is fixed after boot.

\begin{figure}[h]
    \centering
    \includegraphics[width=0.95\linewidth]{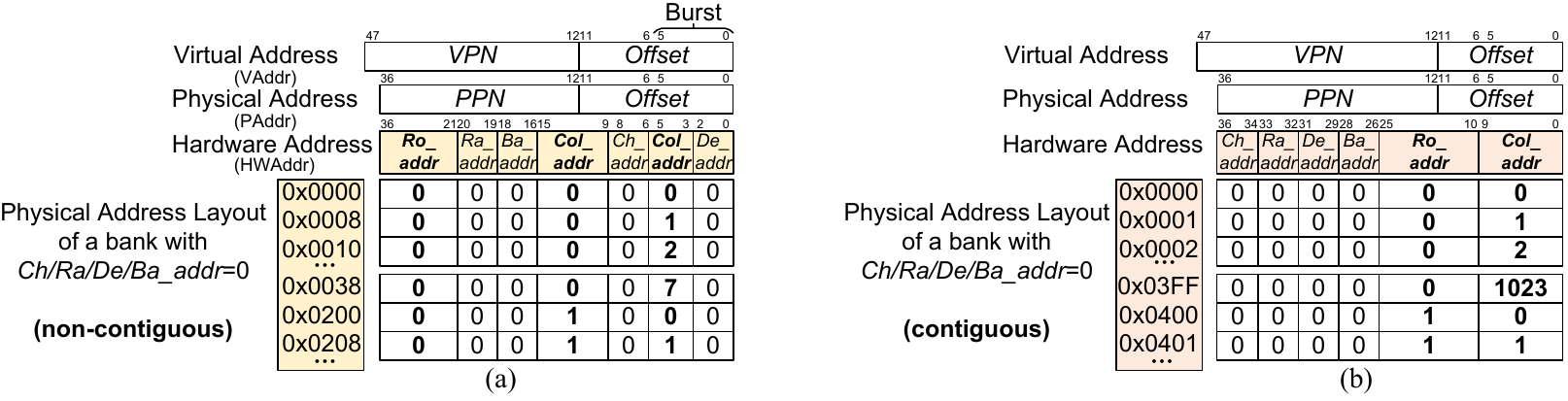}
    \vspace{-10pt}
    \caption{\revisenew{Physical address layout inside one memory bank ($Ch/Ra/Ba/De\_addr$=0) and memory mapping rule for (a) CPU systems with memory interleaving turned on (b) PIM systems with memory interleaving off.}}
    \vspace{-10pt}
    \label{fig:interleave}
    \Description{This figure illustrates the data address arrangement inside one memory bank (ChRaBaDe=0) and the memory mapping rule for (a) CPU systems with global memory interleaving turned on (b) PIM systems with memory interleaving off.}
\end{figure}

Memory interleaving enhances memory bandwidth by mapping adjacent data across different DRAM devices for parallel access.
For example, the DDR4 DRAM hierarchy levels, in descending order of switching overhead, are row, rank, bank, column, and channel \cite{mcjacob2010memory}.
The simplest interleaving scheme maps lower-order address bits to memory hierarchy levels with lower switching overhead, (\figureautorefname{}~\ref{fig:interleave}(a)). 
At the device level, devices (i.e., chips) are grouped to expand data bus width \cite{mcjacob2010memory}.
Individual devices are not separately addressable and share common bank address signal lines within a rank \cite{mcjacob2010memory}.
Therefore, they must be accessed simultaneously, and device-level interleaving cannot be controlled via CPU-side address mapping.
To accommodate various access patterns, practical memory interleaving strategies often employ sophisticated mapping~\cite{hillenbrand2017physical,choi2015study,DRAMA10.5555/3241094.3241139,dramdig_3437539.3437628,one_bit_flip_3241094.3241097}.
Nevertheless, they all depend on the high-order bits of the physical address to select rows due to significant row-switching overhead. 
Additionally, they scatter contiguous data across different banks and devices, which goes directly against good data mapping for PIM systems.

In contrast with CPUs, general-purpose PIM units require the ability to access a contiguous block of data in their own memory.
Therefore, the ideal PIM address mapping maps lower-level memory hierarchy addresses to lower-order bits, as shown in \figureautorefname{}~\ref{fig:interleave}(b), ensuring contiguous physical addresses to reside inside a DRAM device. 
This requirement contradicts the CPUs' memory interleaving.
Current general-purpose PIM systems adopt a software-only approach to emulate the address mapping for PIM units.
They allocate or reserve a large block of contiguous physical addresses as PIM memory space and lay out the data in the PIM memory space with CPU software \cite{UPMEM8875680,UPMEM-Toolchain,PIM-HBM9073325,prim9771457}.
IMPICA \cite{IMPICA7753257} proposes to use huge pages to reduce the address translation overhead in PIM memory space.
However, due to the byte-level memory interleaving in DRAM, these software-only methods introduce significant data re-layout and address mapping overheads.

\subsection{Data Transfer Overhead in DRAM-based PIM}
\label{sec:background:pim}

\begin{figure}[h]
    \centering
    \includegraphics[width=0.95\linewidth]{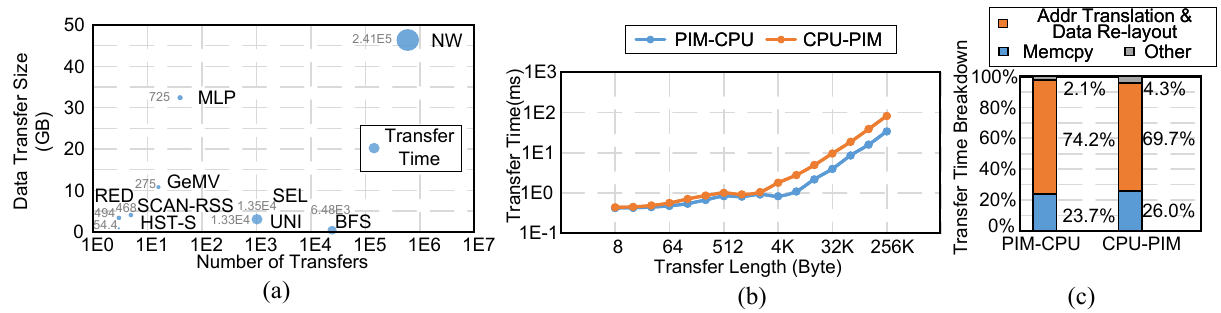}
    \vspace{-10pt}
    \caption{(a) Transfer time of benchmarks. (b) Transfer time across different sizes of a single transfer~(contiguous data). (c) Time breakdown of Data transfer (length = 256KB). Measured with UPMEM SDK \cite{UPMEM-Toolchain}.}
    \vspace{-10pt}
    \label{fig:xfer_breakdown}
    \Description{This figure illustrates the transfer time of benchmarks, the transfer time across different sizes of a single transfer~(contiguous data), and the time breakdown of Data transfer (length = 256KB). Measured with UPMEM SDK \cite{UPMEM-Toolchain}.}
\end{figure}

\revisenew{As we mention in \sectionautorefname{}~\ref{sec:introduction}, disabling memory interleaving preserves data continuity at the rank and channel levels, thereby eliminating their corresponding re-layout overhead \cite{UPMEM8875680,AxDIMM10.1145/3533737.3535093,NDP7429299}.}
However, the memory interleaving of lower-level hierarchies (e.g., device level) cannot be disabled via BIOS, necessitating software to re-layout the data.
\figureautorefname{}~\ref{fig:xfer_breakdown}(c) depicts the breakdown of the data transfer time measured on UPMEM \cite{UPMEM-Toolchain}, whose PIM units are at DDR4's bank level.
Even with channel- and rank-level interleaving disabled, address translation and data re-layout still account for nearly 70\% of the total transfer time. 
\revisenew{UPMEM SDK \cite{UPMEM-Toolchain} provides a multi-threaded API to accelerate the above data transfer.}
However, this introduces a fixed preparation time for each transfer and leads to a non-negligible impact when the length of a single transfer is short.
As shown in \figureautorefname{}~\ref{fig:xfer_breakdown}(b), the transfer time plateaus when the length of a single transfer is less than 4kB.
\figureautorefname{}~\ref{fig:xfer_breakdown}(a) depicts the total transfer size and the number of transfers of different applications.
When the total amount of transferred data is similar, applications with more and shorter transfers, e.g. NW, BFS, and UNI, exhibit a more significant transfer time.
This underscores the need for general-purpose PIM systems to eliminate data transfer between the two memory spaces (i.e. PIM and CPU), and accelerate address mapping and data re-layout.

Li et al. \cite{Trans-FW10071054} compare the address translation overhead of hardware and software methods, finding that the hardware approach reduces overhead by 4.5\texttimes{}.
Therefore, a hardware solution is necessary to reduce the dynamic address mapping overhead in PIM systems.

\subsection{Dynamic Memory Mapping}
\label{sec:background:dynamic}

Dynamic address mapping is widely researched in CPU systems \cite{channel_partition_7851487,DREAM10.1145/2989081.2989102,Valley8416826,Multiple10.1145/3124680.3124742,SoftwareDefined10.1145/3503222.3507774}. DReAM \cite{DREAM10.1145/2989081.2989102} and PAE \cite{Valley8416826} add hardware on the DRAM memory side to dynamically recognize the better-performing address mapping and transparently change the address mapping method.
However, this transparency hides device physical addresses from the host, failing to guarantee the operand co-location required by PIM units.
Multiple physical mappings \cite{Multiple10.1145/3124680.3124742} and software-defined address mapping \cite{SoftwareDefined10.1145/3503222.3507774} enables dynamic mapping by configuring memory controllers and adding CPU-side hardware, respectively.
However, these methods limit interleaving control to levels above the bank level due to memory controller constraints.
Moreover, the method in \cite{SoftwareDefined10.1145/3503222.3507774} modifies the memory management model, requiring changes to the system software as well as hardware.
It is important to enable control over the interleaving of all DRAM hierarchy levels solely via DRAM-side hardware modifications, without requiring changes to system software.

\section{Dual-Track Memory Management}
\label{sec:memory-management}

\revisenew{\archname{}'s dual-track memory management adopts separate management for CPU pages and PIM pages in the unitary, malleable memory space, allowing both to be optimized without compromising the other. 
Global memory interleaving is turned on to preserve CPU memory bandwidth. 
On top of this interleaved memory space, we keep the existing virtual page management and address mapping rules unchanged for CPU pages to preserve compatibility, while proposing chunk-based memory management for PIM pages to enable locality. 
This section presents our approach for bank-level DIMM DRAM PIM systems \cite{UPMEM8875680}, focusing on the address layout and access rules of the chunk-based scheme. 
The hardware implementation of this layout is presented in \sectionautorefname{}~\ref{sec:hw}.
}

\subsection{Chunk-based Memory Management for PIM Pages}
\label{sec:memory-management:chunk}

\revisenew{On the software side of the \archname{} system, the key idea to \emph{prevent the OS from arbitrarily mapping PIM pages in an interleaved memory space is to enforce a uniform distribution of PIM pages across all PIM units' local memory.} 
We thereby present chunk-based address management for PIM pages.}

A \emph{chunk} is a large virtual address block, which is mapped to a contiguous block of physical addresses.
In Linux, the chunk can be allocated through the transparent huge page (THP) mechanism.
As mentioned in \sectionautorefname{}~\ref{sec:background:interleave}, most interleaving strategies map the Row address to the highest-order bits of the physical address.
\revisenew{
\emph{Based on this observation, if the chunk size ($S_C$) is large enough~($S_C \geq \text{\#Channels} \times \text{\#Ranks} \times \text{\#Banks} 
\times \text{\#Devices} \times \text{row buffer size}$), with the global memory interleaving turned on, the chunk is evenly distributed across all memory banks.} For example, the memory system configuration shown in \figureautorefname{}~\ref{fig:interleave} comprises 8 channels, 4 ranks per channel, 8 banks, and 8 devices per rank.
}
If the chunk size is set to 256~MB, each bank would contribute $S_B = S_C / (8 \times 4 \times 8 \times 8)$ = 128kB of data, i.e., the PIM page size. This means that when a PIM chunk is allocated, each PIM unit gets a 128kB \emph{PIM page} in its PIM local memory.

\subsection{CPU Addressing PIM Pages}
\label{sec:memory-management:cpu-addressing}

\begin{figure}[h]
    \begin{minipage}{0.58\linewidth}
        \centering
        \includegraphics[width=\linewidth]{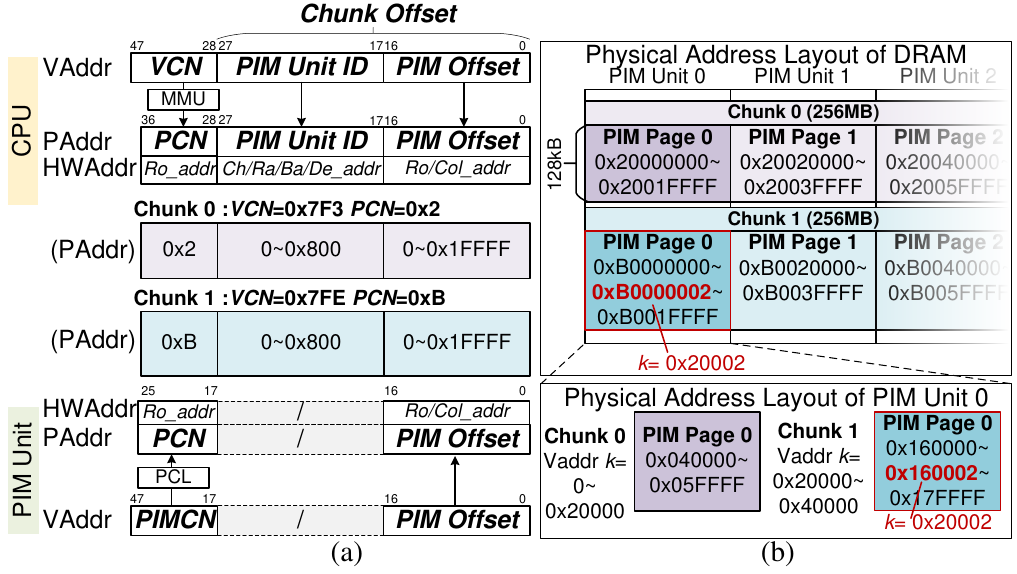}
    \vspace{-20pt}
        \caption{\revisenew{Chunk-based management for PIM pages. (a) Address mapping. (b) Data layout.}}
    \vspace{-10pt}
        \label{fig:allocation}
        \Description{This figure illustrates the chunk-based management for PIM pages. (a) Address mapping. (b) Data layout.}
    \end{minipage}
    \hfill
    \begin{minipage}{0.38\linewidth}
        \centering
        \includegraphics[width=\linewidth]{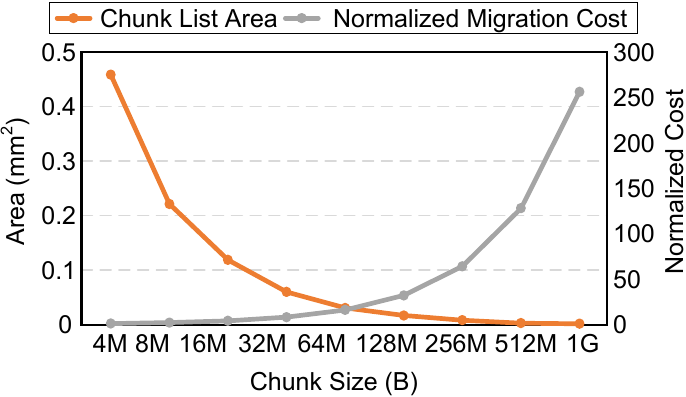}
    \vspace{-10pt}
        \caption{ Page migration cost and chunk list area under different chunk sizes. }
    \vspace{-10pt}
        \label{fig:chunk_size}
        \Description{This figure illustrates the page migration cost and chunk list area under different chunk sizes.}
    \end{minipage}
\end{figure}

\revisenew{A preferred address layout inside PIM chunks is shown in \figureautorefname{}~\ref{fig:allocation}(b), where adjacent addresses are preferentially mapped to the same PIM unit.
\figureautorefname{}~\ref{fig:allocation}(a) shows its corresponding address mapping, including the composition of the virtual, physical, and hardware addresses and their translation and mapping.
VAddr[47:28] serves as the \emph{virtual chunk number (VCN)} to select the chunk and is translated to the \emph{physical chunk number (PCN)} that is used for recognizing PIM pages.
VAddr[27:0] is the \emph{chunk offset} and is passed unchanged to PAddr[27:0] during address translation.
PAddr[27:17] in the chunk offset are mapped sequentially to channel, rank, bank, and device, encoding the unique identifier of the target PIM unit (\emph{PIM unit ID}) for software tracking.
Finally, PAddr[16:0] are mapped to row and column inside the 128kB PIM pages, referred to as \emph{PIM offset}.}

\revisenew{
In practice, one 128kB PIM page for each PIM unit might not be enough for PIM tasks, so multiple chunks are allocated.
The CPU host program records the virtual chunk numbers of these chunks in an array ($VCNs$).
Given the above layout, when the host CPU needs to compute the virtual address of a data byte in a PIM unit's local memory, it must derive the byte's location within the chunk's virtual address space.
Using \figureautorefname{}~\ref{fig:allocation}(a) as an example, two 256MB chunks ($S_C$=256MB) are allocated.
The two chunks (with VCN=\texttt{0x7f3} and \texttt{0x7fE}) are mapped to physical addresses with PAddr[36:28]= \texttt{0x2} and \texttt{0xB}, respectively.
Suppose the CPU host program needs to locate the $k^\mathrm{th}$ byte~(e.g., $k$=\texttt{0x20002}) within the concatenated data of all the PIM pages in PIM unit 0 ($i=0$); the CPU can first derive which chunk the address belongs to ($N$) and the PIM offset ($off$) by:}
\begin{eqnarray}
    N = k / S_B = 1, off = k \bmod S_B=2
    \label{eqn:chunk_num}
\end{eqnarray}
and then computes the virtual address of this byte by:
\begin{eqnarray}
VAddr & = & VCNs[N] << 28 + i \times S_B + off = \texttt{0x7fE0000002}
\end{eqnarray}
This virtual address is further translated to the physical address \texttt{0xB0000002} by the MMU and mapped to the hardware address by address mapping modules.

\subsection{PIM Units Addressing PIM Pages}
\label{sec:memory-management:pim-addressing}

For memory isolation and correctness, each PIM unit must access only its designated PIM pages and be prevented from accessing CPU pages. 
However, since \archname{} relies on the host OS to allocate PIM pages from the CPU side, the PIM units are inherently unaware of these physical addresses at runtime.

To address this, we introduce a dedicated virtual address space for PIM units and synchronize the physical chunk number information from the host CPU.
In our example (\figureautorefname{}~\ref{fig:allocation}), the two allocated chunks provide a total capacity of 256kB ($2 \cdot S_B$) for each PIM unit's virtual and physical address spaces.
\figureautorefname{}~\ref{fig:allocation}(a) depicts the correspondence between PIM's virtual address and hardware address.
Notably, unlike the CPU's address mapping, a PIM unit's address does not contain its own PIM unit ID.
Instead, VAddr[47:17] indicates the PIM-side chunk number (PIMCN). 

Suppose PIM unit $i=0$ accesses the same byte (the $k^\mathrm{th}$ byte, $k$=\texttt{0x20002}) as in our example in \sectionautorefname{}~\ref{sec:memory-management:cpu-addressing}; it generates $k$ as the virtual address.
Then the PIM chunk number is $N=1$.
The PIM chunk number is translated to the physical chunk number by the address translation circuit, i.e., the PIM Chunk List (PCL), which is presented in \sectionautorefname{}~\ref{sec:hw:pim_translate}.
The physical chunk number of the PIM side is the same as that of the CPU side (=\texttt{0xB}).
With the same address mapping rule, they point to the same hardware address.
Bits 16-0 of both the virtual address and the physical address represent the PIM offset ($off=2$), which equals that of the CPU side.
Finally, the physical chunk number and PIM offset are combined into the physical address \texttt{0x160002} and mapped to the same hardware address as the CPU side.

\subsection{Chunk Size Selection}
\label{sec:app:chunk_size}

Chunk size is a critical system parameter. 
Generally, allocating a huge page larger than 4MB may require page migration to accommodate it.
\revisenew{A larger chunk size inherently encompasses more occupied pages, thereby increasing the number of pages that must be migrated.}
Conversely, a small chunk size necessitates allocating more chunks to satisfy PIM units' capacity demands.
This in turn requires a larger PIM Chunk List to store PIM chunk numbers, increasing storage overhead.
\figureautorefname{}~\ref{fig:chunk_size} displays the migration overhead and PIM Chunk List area across varying chunk sizes.
Critically, since PIM memory allocations are typically long-lived and reused throughout the application execution, multiple runtime allocations and releases of PIM chunks can be largely avoided.
Therefore, we prioritize minimizing the PIM Chunk List area and select 256~MB as the chunk size.
In practice, this size can be adapted to any standard huge page configuration supported by the host OS.

\section{Hardware Support for Page Layouts}
\label{sec:hw}

\begin{figure}[h]
    \centering
    \includegraphics[width=\linewidth]{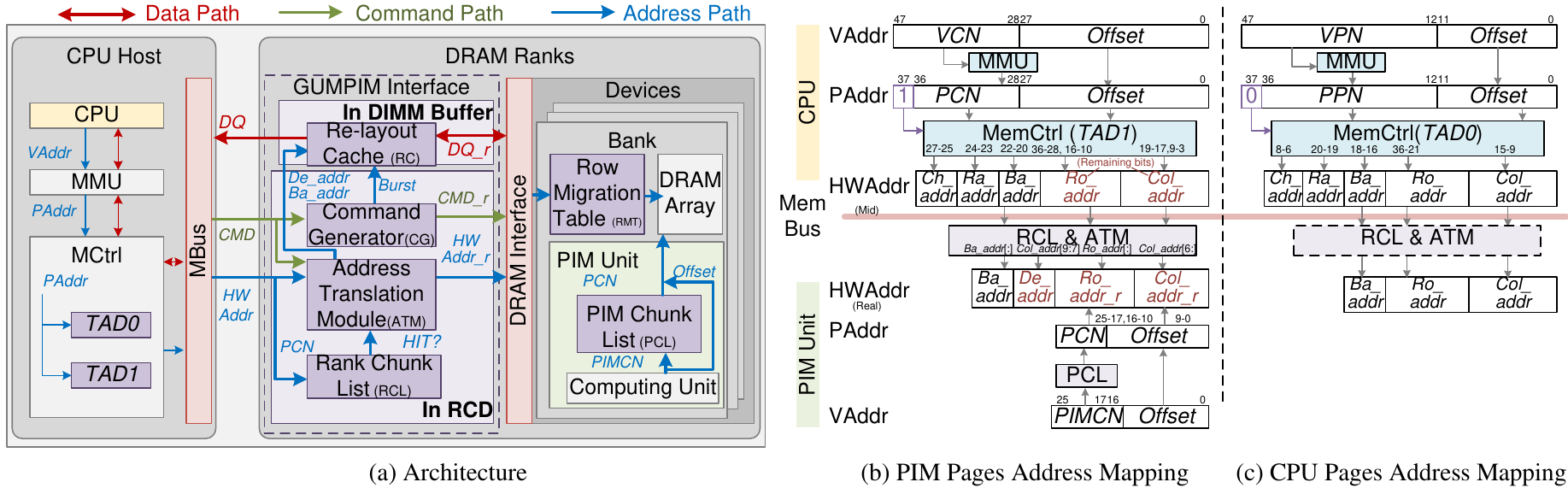}
    \vspace{-20pt}
    \caption{(a) \archname{} architecture. (b)(c) Address mapping path for PIM and CPU pages.}
    \vspace{-10pt}
    \label{fig:arch_mapping}
    \Description{This figure illustrates the \archname{} architecture and the address mapping path for PIM and CPU pages.}
\end{figure}

\revisenew{
This section details \archname{}'s hardware design supporting the chunk-based address layout presented in \sectionautorefname{}~\ref{sec:memory-management}, using DIMM DRAM \cite{UPMEM8875680}, whose hardware architecture comprises three layers, as shown in \figureautorefname{}~\ref{fig:arch_mapping}(a). 
On the CPU side, we configure the memory controller to apply distinct address mappings for PIM and CPU pages at and above the bank level, requiring no hardware modification.
In each DIMM, we insert the \archname{} interface, which integrates hardware modules into the RCD and DIMM buffer chips \cite{AxDIMM10.1145/3533737.3535093} to handle address mapping below the bank level and data re-layout for PIM pages. 
Inside each bank, we introduce a hardware module for PIM-side address translation. 
The adaptation to other DRAM types (LPDDRx and HBMx) is presented in \sectionautorefname{}~\ref{sec:other_dram}.
}

\subsection{CPU-side Address Mapping}
\label{sec:hw:mapping}

As mentioned in \sectionautorefname{}~\ref{sec:memory-management}, CPU and PIM pages have different data layouts and require dynamic address mapping.
For CPU pages, the address mapping path is kept unchanged as configured in the BIOS.
The address mapping for a PIM page involves two steps: above-bank and below-bank address mapping.
Above-bank address mapping is implemented by configuring the CPU's BIOS.
For below-bank address mapping, hardware modules are introduced on the DRAM side.
This is because the intra-rank device organization is entirely transparent to the host CPU and cannot be configured via BIOS.

\subsubsection{Above-Bank Address Mapping}
\label{sec:hw:mapping:inter}

For address mapping of hierarchies at and above the bank level (i.e., channel, rank, and bank in DIMM DRAM), we adopt the multiple address mappings technique \cite{Multiple10.1145/3124680.3124742}.
With this approach, an address alias with a different address mapping is added by configuring the memory controllers through the BIOS.
For example, in the Xeon processor, an alias can be implemented by appending a TAD region, $TAD1$, to the PCI configuration space.
$TAD1$ points to the same memory space as the default $TAD0$ (\figureautorefname{}~\ref{fig:arch_mapping}(c)), but with a non-interleaved address mapping, as shown in \figureautorefname{}~\ref{fig:arch_mapping}(b).

In our example with a 256GB memory system, the interleaved physical address space spans [0, 256GB), while the non-interleaved address alias spans [256GB, 512GB).
Bit 37 of the physical address determines which TAD region an address belongs to.
If bit 37 is \texttt{0}, the address is from a CPU page, using $TAD0$ for address mapping.
Otherwise, the address is from a PIM page and $TAD1$ is used.
After determining the used TAD, the memory controller performs partial address mapping.
In this step, only physical address bits for rank, channel, and bank are mapped to the real hardware address.
The remaining bits are transferred to the DRAM side \archname{} interface for below-bank address mapping.
This means that the hardware address generated by the memory controller at this step is an intermediate result and is to be further mapped by the \archname{} interface presented in \sectionautorefname{}~\ref{sec:hw:mapping:intra}.
To comply with the DRAM protocol, the remaining bits are divided into row and column addresses ($Col\_addr$ and $Ro\_addr$ in \figureautorefname{}~\ref{fig:arch_mapping}(b)) and sent to DRAM.

\subsubsection{Below-Bank Address Mapping}
\label{sec:hw:mapping:intra}
Dynamic address mapping of levels below the bank (i.e., row, device, and column), as illustrated in \figureautorefname{}~\ref{fig:arch_mapping}(b), is carried out by three main hardware modules, the Rank Chunk List (RCL), Address Translation Module (ATM), and Command Generator (CG), in the \archname{} interface.
The three hardware modules are integrated into the RCD chip, as shown in \figureautorefname{}~\ref{fig:arch_mapping}(a).
The Rank Chunk List determines the page type, while the Address Translation Module maps the remaining bits of the intermediate hardware address to row, column, and device addresses.
Command Generator (CG) generates a command with the real address for the DRAM interface.

\textbf{Rank Chunk List}.
Each row of the Rank Chunk List stores the physical chunk number of one chunk. Upon receiving a row address, the highest 9 bits of the address are matched against the list. A hit (HIT=1) indicates a PIM page, while a miss (HIT=0) indicates a CPU page.

\textbf{Address Translation Module and Command Generator}.
As depicted in \figureautorefname{}~\ref{fig:arch_mapping}(c), if the Rank Chunk List detects that the address belongs to CPU pages, the Address Translation Module keeps the row and column addresses unchanged.
Otherwise, the Address Translation Module maps bits 9-7 of the original column address ($Col\_addr$) as the device address ($De\_addr$), and other bits are left-shifted by 3 (zero-padded at the least significant bits) to generate the real column address ($Col\_addr\_r$).
The row and column addresses are sent to the DRAM interface for data access, and the device and bank addresses are sent to the Re-layout Cache (RC) for further data selection.
We present the procedure in \sectionautorefname{}~\ref{sec:hw:RC}.
After mapping the address, the Command Generator generates the real command for the DRAM interface.

\textbf{Timing of Address Mapping.}
Although the \archname{} interface is in the critical path between the memory controller and the DRAM interface, it preserves native DRAM device timings.
\figureautorefname{}~\ref{fig:timing}(a) depicts a read (\texttt{RD}) request and its prerequisite requests, precharge (\texttt{PRE}) and activate (\texttt{ACT}).
After receiving a precharge request, the \archname{} interface starts to convert the address and then sends the real precharge request to the DRAM interface.
The memory bus, \archname{} interface, and DRAM interface form a pipeline.
As a result, the DRAM timing $tRP$ and $tRCD$ are kept unchanged.
In effect, only a fixed delay $tUI$ is incurred on end-to-end read latency without accumulating across pipelined commands, so its impact on the overall read latency is minimal.

\subsection{PIM-side Address Translation}
\label{sec:hw:pim_translate}

\textbf{PIM Chunk List.}
The PIM Chunk List contains the physical chunk numbers allocated by the CPU, where each entry is indexed by its corresponding PIM chunk number ($N$).
When the PIM unit generates a virtual address, the PIM Chunk List looks up the $N^\mathrm{th}$ row and gets the corresponding physical chunk number.
The physical chunk number is further combined with the PIM offset to obtain the physical address.

\subsection{Data Re-layout}
\label{sec:hw:RC}

\begin{figure}[h]
    \centering
    \includegraphics[width=\linewidth]{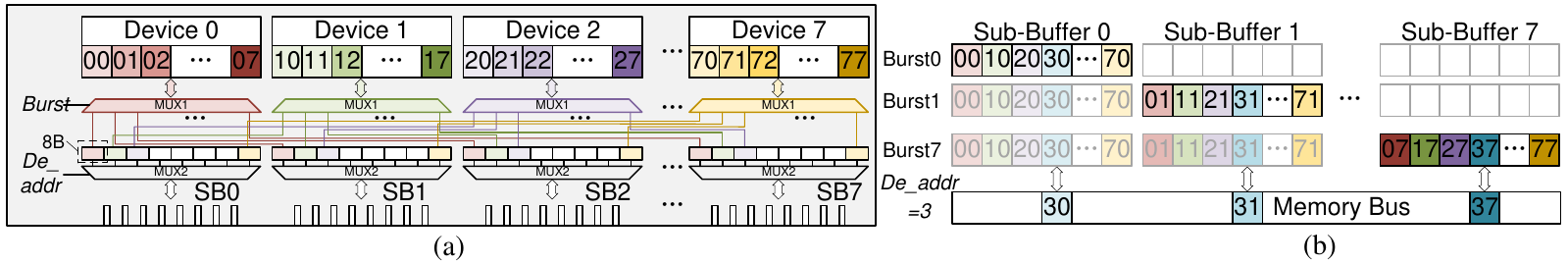}
    \vspace{-20pt}
    \caption{(a) RC-to-device connection. (b) Data re-layout with Re-layout Cache.}
    \vspace{-5pt}
    \label{fig:rc}
    \Description{This figure illustrates the Re-layout Cache-to-Device connection and data re-layout with the Re-layout Cache.}
\end{figure}

The DRAM is designed to access data at burst granularity \cite{JEDEC_JESD238A_2020_ddr4,JEDEC_JESD238A_2020_lpddr5,JEDEC_JESD238A_2023_hbm3}.
Assume a DRAM burst size of 64B and a bit width of 64 bits.
In each burst, the DRAM interface accesses 8B of data from every device to fully utilize the device-level parallelism.
For example, as illustrated in \figureautorefname{}~\ref{fig:rc}, one burst consists of data \texttt{00}, \texttt{10}, \dots, \texttt{70}, which are distributed across all devices of a rank.
However, according to the mapping rules of PIM pages, a consecutive 64B range of a PIM page, denoted as \emph{PIM burst}, is mapped to the same device.
For instance, one PIM burst comprises data \texttt{00}, \texttt{01}, \dots, \texttt{07}. 
Accessing such a complete 64B range requires 8 normal burst accesses, with each device providing 64B of data. 
The remaining $7 \times 64$B, however, remain unused, resulting in a waste of memory bus bandwidth and requiring CPU intervention to filter out these unused cache lines.
Therefore, we propose adding a Re-layout Cache module to buffer, re-layout, and assemble the contiguous 64B on the DRAM side, conserving memory bus bandwidth and mitigating the need for CPU involvement.

\begin{figure}[h]
    \centering
    \includegraphics[width=0.85\linewidth]{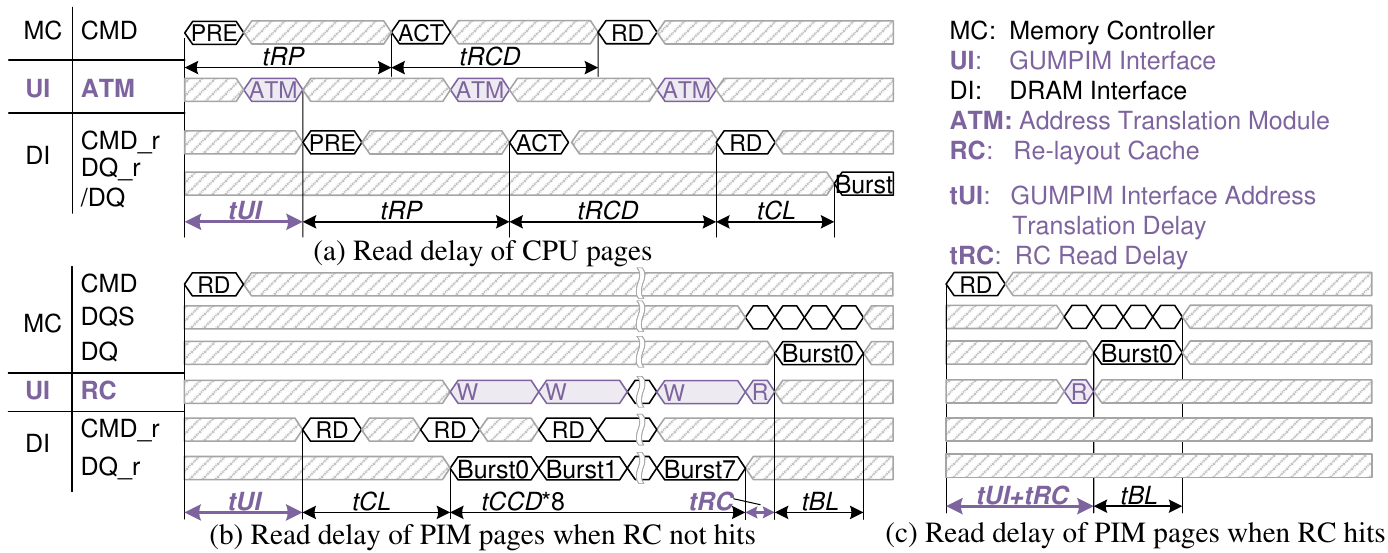}
    \vspace{-10pt}
    \caption{Timing of \archname{} interface. }
    \vspace{-10pt}
    \label{fig:timing}
    \Description{This figure illustrates the timing of \archname{} interface.}
\end{figure}

\textbf{Re-layout Cache}.
The Re-layout Cache circuits are placed between the memory bus and the data signal line of the DRAM interface and are integrated into the DIMM buffer \cite{AxDIMM10.1145/3533737.3535093}.
With a burst size of 64B and a bit width of 64, each Re-layout Cache contains eight sub-buffers (SBs), each consisting of eight cells that store 8B of data, as depicted in \figureautorefname{}~\ref{fig:rc}(a).
\figureautorefname{}~\ref{fig:rc}(b) shows an example of reading a PIM burst from the third DRAM device ($De\_addr$=3), and \figureautorefname{}~\ref{fig:timing}(b) depicts the timing.
The Address Translation Module converts the column address ($Col\_addr$) from the memory controller into the real column address ($Col\_addr\_r$).
Note that the lower 3 bits of $Col\_addr\_r$ are zero, as mentioned in \sectionautorefname{}~\ref{sec:hw:mapping:intra}.
The Command Generator then generates 8 burst access commands ranging from column $Col\_addr\_r$ to $Col\_addr\_r+7$. 
At the end of each burst, the MUX1 of every sub-buffer is selected by the $Burst$ signal generated by the Command Generator, i.e., the $i$\textsuperscript{th} ($i=0,\dots,7$) burst fills the $i$\textsuperscript{th} sub-buffer.
After the 8 bursts, all the sub-buffers are filled.
Then the $De\_addr$ signal (=3) is used to select MUX2, and the 3\textsuperscript{rd} cell of each SB is pushed to the memory bus and transferred to the CPU side.

The Re-layout Cache enables fast data re-layout and filtering on the DRAM side without CPU participation.
However, it does not prevent DRAM banks from reading redundant bytes.
We adopt a delayed update scheme to improve the chances of using the redundant bytes already fetched into the Re-layout Cache, i.e., updating only on a Re-layout Cache miss.
\figureautorefname{}~\ref{fig:timing}(c) shows the timing when data has already been fetched into the Re-layout Cache.
When accessing data from PIM chunks, the Command Generator first checks whether the data has already been fetched into the Re-layout Cache by comparing the row, bank, and column addresses.
If the data has already been fetched, the Command Generator can directly initiate the burst transfer by reading from the Re-layout Cache, which saves $tCL$ time.
Therefore, to maximize performance, the compiler should explicitly optimize memory access patterns to improve the Re-layout Cache hit rate.

\subsection{DRAM Instruction Support for Dual-Track Memory Management}
\label{sec:hw:ins}

We summarize the instructions that support dual-track memory management in \tablename{}~\ref{tab:instruction}.
These instructions are implemented via MMIO operations exposed as normal memory read/write or register configurations \cite{UPMEM-Toolchain,UPMEM8875680} rather than extending the CPU ISA or DRAM standard commands.
Consequently, the instruction types, hierarchical organization, and state machine rules of the DRAM interface remain completely unchanged.
Specifically, the \texttt{ACN} instruction appends an item to the PIM Chunk List and Rank Chunk List when a new PIM chunk is allocated, while the \texttt{CCL} instruction clears the chunk lists upon program termination.
Additionally, the \texttt{CRC} instruction flushes data in the Re-layout Cache to DRAM banks to avoid data inconsistency caused by the delayed update scheme.

\begin{table}[h]
    \centering
    \caption{DRAM-PIM Instruction Extension.}
    \vspace{-10pt}
    \label{tab:instruction}
\resizebox{0.7\linewidth}{!}{%
    \begin{tabular}{c|c|c|c}
        \toprule
        Technique   & Command & Parameter   & Description               \\ \midrule
        Dual-Track      & ACN     & Chunk No.   & Append chunk number to RCL \& PCL  \\
        Memory Management  & CCL     & -           & Clear RCL \& PCL                    \\
        & CRC     & -           & Write data in RC back to DRAM banks \\
                    \midrule
        Guaranteed PIM   & SRMT    & Srcs \& Dests & Append (Srcs, Dests) row pairs to RMT \\
        Chunk Allocation & CRMT    & -           & Clear RMT \\ \bottomrule
    \end{tabular}
}
    \vspace{-10pt}
\end{table}

\section{Software Support for \archname{}}
\label{sec:api}

In this section, we provide API support for memory management and specific memory access modes.
We also provide APIs for specific data transfer operations between PIM units, i.e., scatter, broadcast, and gather.

\subsection{Memory Allocation}

For CPU pages, we keep the \texttt{malloc()} function in glibc unchanged.
For PIM pages, we define an API function \texttt{malloc\_pim($len$)} to allocate memory with length $len$ for all PIM units.
The API calculates the required chunk count based on $len$, allocates chunks through the \texttt{mmap()} system call, and marks them as THPs through \texttt{madvise()}.
Due to demand paging, the CPU must access at least one byte per chunk to trigger physical page allocation.
Since PIM units cannot handle page faults, the \texttt{mlock()} system call is invoked to prevent PIM chunks from being swapped out.
Finally, the API queries \texttt{/proc/pid/pagemap} for the physical address and extracts the physical chunk number, which is then registered to the Rank Chunk List and PIM Chunk List via the \texttt{ACN} instruction.
Upon program termination, the \texttt{CCL} instruction clears the two Chunk Lists.

\subsection{Unit-Aware Memory Access}

To simplify memory access across multiple chunks within a PIM unit, we provide unit-aware APIs that allow developers to access data by specifying the PIM unit ID and PIM offset, without manually computing virtual addresses. 
The unit-aware APIs are summarized as follows:
\begin{itemize}
    \item \texttt{pim\_get\_ptr($unit\_id$, $offset$)} returns a pointer to the memory at the specified offset within the given PIM unit's local memory, facilitating random access.
    \item \texttt{pim\_copy($dst\_unit\_id$, $dst\_offset$, $src\_unit\_id$, $src\_offset$, $length$, $direction$)} handles memory copy operations with automatic cross-chunk boundary management, where $direction$ specifies the transfer direction: \texttt{PIM\_TO\_CPU}, \texttt{CPU\_TO\_PIM}, or \texttt{PIM\_TO\_PIM}.
    \item \texttt{pim\_iterator($unit\_id\_list$, $start\_off$, $end\_off$)} provides an iterator interface to traverse consecutive data within one or multiple PIM units across chunk boundaries.
\end{itemize}
These APIs provide a linear address space abstraction for each PIM unit, automatically handling the wrap-around at chunk boundaries and the crossing of chunk boundaries within a PIM unit's linear address space.
This enables straightforward operations like vector traversal and multi-operand alignment without the need for manual and error-prone address computation.

\subsection{Performance Impact of Different Nested Loop Order}

After PIM computation, CPUs must traverse the results for further processing or inter-PIM communication.
The traversal order significantly impacts the Re-layout Cache hit rate and DRAM bandwidth.
For instance, sequential traversal of a single device (with 8-way device interleaving) reads 8\texttimes{} redundant data from DRAM. 
Although the Re-layout Cache filters this redundancy before it reaches the memory bus, DRAM bandwidth utilization remains only 1/8.

\begin{figure}[h]
    \centering
    \includegraphics[width=\linewidth]{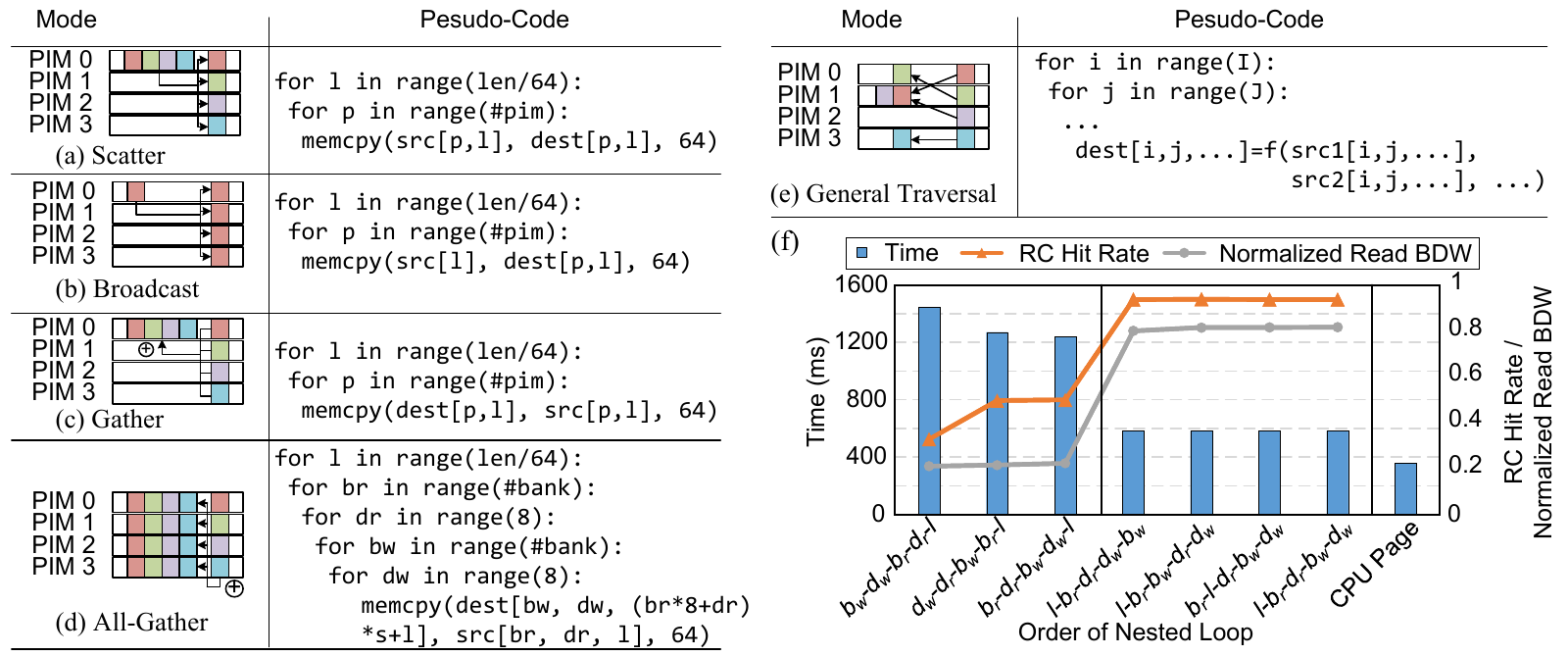}
    \vspace{-10pt}
    \caption{(a-e) Inter-PIM-unit communication and CPU traversal on the results of the PIM computation. (f) Time, Re-layout Cache hit rate, and read bandwidth (normalized to CPU Page) of all-gather on 8 PIM ranks with different nested-loop orders. The data block size is 1024. Left: the order disobeys at least one rule. Middle: the order obeys all the rules. Right: an all-gather operation on CPU pages.}
    \vspace{-10pt}
    \label{fig:access_mode_and_loop_order}
    \Description{}
\end{figure}

For example, in all-gather communication, which reads data from all PIM units and broadcasts it back, the loop structure (\figureautorefname{}~\ref{fig:access_mode_and_loop_order}(d)) contains five nested loops: offset inside each PIM unit's bank $l$, read/write bank addresses $b_r/b_w$, and read/write device addresses $d_r/d_w$. 
To maximize Re-layout Cache hit rate, device address loops ($d_r$ and $d_w$) must be innermost. Specifically:
(1) For reads, $b_r$ and $l$ should be outside $d_r$;
(2) For writes, $l$, $b_r$, and $d_r$ should be outside $d_w$.
\figureautorefname{}~\ref{fig:access_mode_and_loop_order}(f) shows all-gather performance under different loop orders. Orders obeying these rules achieve higher Re-layout Cache hit rates and DRAM bandwidth, resulting in performance close to the CPU-page baseline (only 1.6\texttimes{} slower than the baseline). 
This optimization strategy also applies to general traversal operations (\figureautorefname{}~\ref{fig:access_mode_and_loop_order}(e)).

\subsection{High-level APIs for Inter-PIM-Unit Data Transfer}

We provide several high-level APIs for widely used inter-PIM-unit communication operations whose performance is significantly affected by the Re-layout Cache hit rate.
Due to the significant cost of communication between PIM units, existing PIM systems often adopt a fork-join computing mode \cite{UPMEM8875680,PIM-HBM9073325,AxDIMM10.1145/3533737.3535093,prim9771457}.
therefore, we adopt communication modes from NCCL \cite{NCCL} and present the nested-loop order for these modes to better utilize the Re-layout Cache.

\textbf{Scatter.}
Scatter reads a contiguous data block from a single DRAM bank and distributes its subblocks across all banks. 
\figureautorefname{}~\ref{fig:access_mode_and_loop_order}(a) presents a scatter that involves 4 PIM units.
In each iteration of the outermost loop $l$, a burst length of data is read from each bank and written to the destination PIM bank.
The Re-layout Cache in each rank buffers data from 8 devices and performs batch writes, thereby fully utilizing its locality.

\textbf{Broadcast.}
Broadcast reads a data block from a single DRAM bank and writes it to all banks.
\figureautorefname{}~\ref{fig:access_mode_and_loop_order}(b) depicts a broadcast originating from PIM 0.
In each iteration of the outermost loop $l$, a 64B block is read from PIM 0 and written to every PIM bank.
This order fully leverages the locality of the Re-layout Cache.

\textbf{Gather.}
Gather is the reverse of scatter, reading data blocks from all DRAM banks and consolidating them into a single bank.
As shown in \figureautorefname{}~\ref{fig:access_mode_and_loop_order}(c), PIM 0 gathers data from every bank. 
The outermost loop $l$ iterates over data block offsets, similar to scatter. 
By reading data with the same offset $l$ in adjacent iterations, the locality of the Re-layout Cache is fully utilized.

\section{Guaranteed PIM Chunk Allocation}
\label{sec:chunk_allocation}

\subsection{Memory Fragmentation and Chunk Allocation}
\label{sec:chunk_allocation:fragmentation}

As presented in \sectionautorefname{}~\ref{sec:memory-management:chunk}, chunk-based PIM page management requires allocating a large block of contiguous memory space.
However, allocating a chunk is challenging in modern operating systems due to severe physical memory fragmentation.
Our experiments on a 256GB memory system show that even at only 25\% memory occupancy, allocating a single 256MB chunk can fail.
Although the OS attempts to form contiguous space via page migration, this approach falls short for two reasons:
First, page migration overhead (\textasciitilde4$\mu$s per page \cite{Contiguitas10.1145/3579371.3589079}) can offset PIM's performance gains.
Second, unmovable pages (e.g., slab and DMA pages that have strict access latency constraints) are scattered throughout the memory, preventing the OS from consolidating contiguous physical memory. Consequently, chunk allocation fails and falls back to normal 4kB pages, violating the contiguous memory requirement of PIM.
Static memory reservation (e.g., hugetlbfs \cite{linux_hugetlbfs_reserv_v5_1}) avoids migration but lacks flexibility and wastes memory for CPU applications.
Thus, overcoming the latency and mobility bottlenecks of dynamic page migration offers a far more practical path to guaranteed chunk allocation.

\subsection{Challenges in CPU-Assisted Page Migration}
\label{sec:chunk_allocation:migration_challenges}

Page migration overhead fundamentally stems from three components: Translation Lookaside Buffer (TLB) shootdown, Page Table Entry (PTE) updates, and data copying.
To reduce these costs, state-of-the-art CPU-assisted designs like Contiguitas \cite{Contiguitas10.1145/3579371.3589079} introduce hardware extensions in the Last-Level Cache (LLC) to buffer migrating pages, allowing host CPUs to access them during migration.
This method reduces the page migration overhead by 2\texttimes{} by eliminating the need for locking and TLB shootdown, and enables migration of otherwise unmovable pages.
Despite these improvements, this cache-based approach introduces two critical limitations.
First, the migrating page copy is buffered in the LLC, forcing I/O devices to redirect accesses to the CPU-side cache and requiring substantial CPU-side hardware modifications.
Second, because data must still be transferred between main memory and the CPU, migration overhead remains high.
The memory copy accounts for 58\% of the total migration overhead (TLB shootdown is excluded) according to our simulation.

\subsection{PIM-Assisted Page Migration}
\label{sec:chunk_allocation:migration}

PIM architectures offer a promising alternative to fundamentally alleviate migration bottlenecks and achieve flexible, on-demand chunk allocation. 
By offloading transparent page migration entirely to custom hardware components on the DRAM side, we can eliminate expensive bus-crossing data movements while keeping the migration process nearly transparent to the host CPU.

The core insight of PIM-assisted page migration is to introduce an address redirection scheme in DRAM, coupled with data movement performed directly by the PIM units. 
This ensures that CPUs and I/O devices access the correct data throughout the entire migration process without software intervention or memory coherence violations.
We deliberately constrain all hardware modifications to the DRAM side, which is architecturally principled: DRAM serves as the ultimate memory layer that is physically shared and directly accessible via physical addresses by all system agents (e.g., CPUs, GPUs, and I/O controllers).
Implementing address redirection directly inside DRAM leverages this universal visibility, enabling seamless, transparent access across all system agents throughout the migration process.

\begin{figure}[h]
    \centering
    \includegraphics[width=\linewidth]{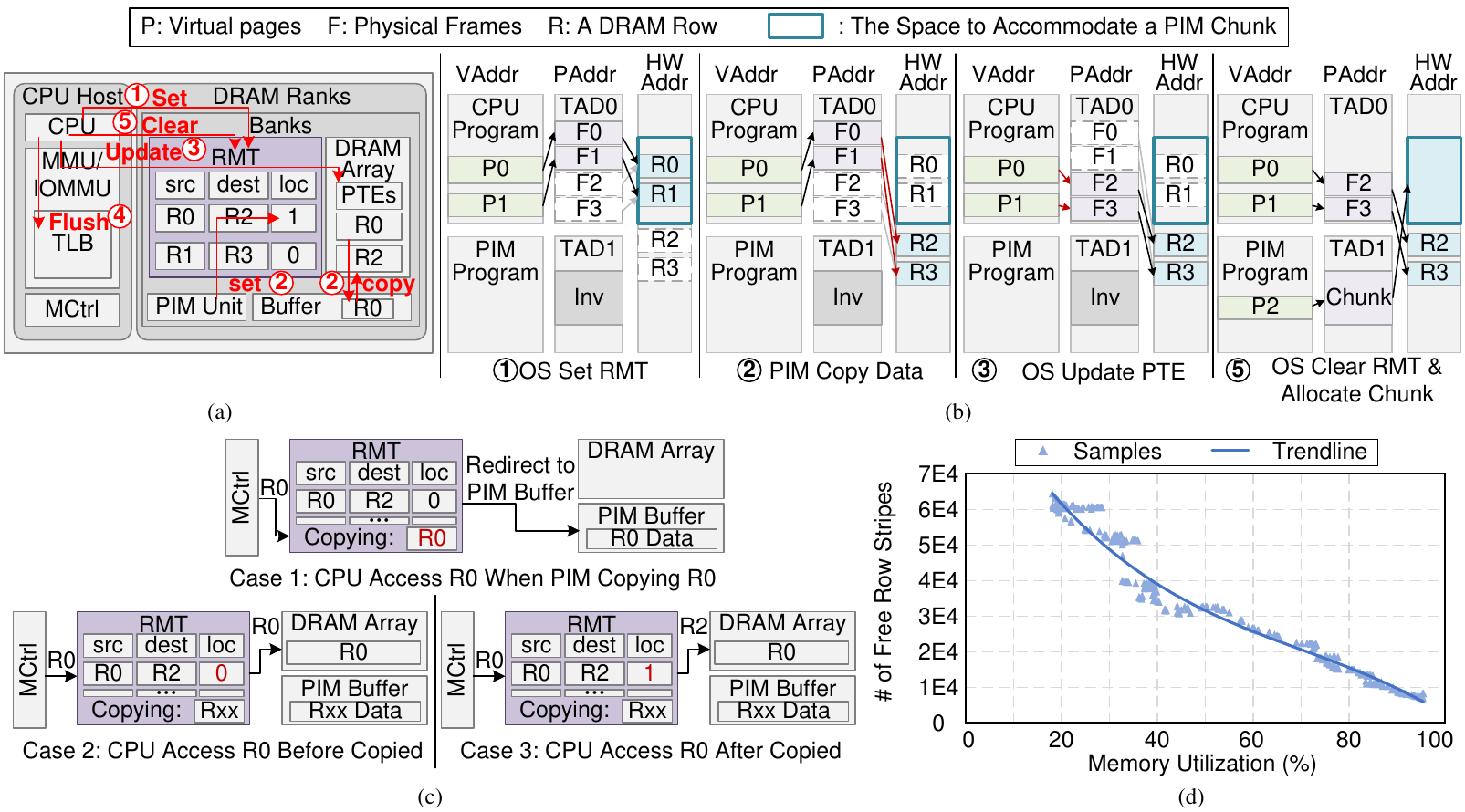}
    \vspace{-20pt}
    \caption{(a) (b) PIM-assisted page migration during chunk allocation. (c) Cases of CPU access to the migrating page. (d) Row stripe count under different memory occupancy. $F0$ denotes the set of 4kB physical page frames whose backing storage is contained within a DRAM row $R0$, since a typical DRAM row spans multiple page frames. $P0$ denotes the set of 4kB virtual pages referencing $F0$.}
    \vspace{-5pt}
    \label{fig:migration}
    \Description{This figure illustrates PIM-assisted page migration during memory allocation.}
\end{figure}

\textbf{Row Migration Table (RMT).}
The Row Migration Table is a lightweight hardware component that resides in each memory bank and tracks data location and migration status, enabling row address redirection throughout migration (\figureautorefname{}~\ref{fig:migration}(a)). 
Each Row Migration Table entry contains two key fields: (1) a location field (\texttt{loc}) indicating whether the row has been physically migrated, and (2) a \texttt{copying} field identifying the row currently being migrated. 
\figureautorefname{}~\ref{fig:migration}(c) illustrates the redirection principle for three cases, covering all possible situations during migration.
In case 1, a PIM unit is copying $R0$ (\texttt{copying} = $R0$).
The data being copied is loaded into the PIM buffer, and the Row Migration Table redirects the access to the PIM buffer if the CPU accesses $R0$.
In case 2, when the CPU accesses $R0$ while the data is still in the source row (\texttt{loc} = \texttt{0}), the Row Migration Table redirects the access to the source row.
In case 3, the migration of $R0$ completes (\texttt{loc} = \texttt{1}), and the Row Migration Table routes the access to the new physical row $R2$.
This hardware-supported redirection ensures the correct data access during migration, maintaining coherence for CPUs and I/O devices until the OS updates the page tables.

\textbf{Migration Workflow.}
When the PIM program allocates a chunk, the OS selects a candidate physical region.
The OS identifies the occupied pages within this region and allocates an equivalent capacity of free physical pages via native memory allocation mechanisms (e.g., the Linux Buddy System \cite{buddy}) to accommodate these pages.
\figureautorefname{}~\ref{fig:migration}(b) illustrates the subsequent page migration process.
The OS issues the \texttt{SRMT} instruction (summarized in \tableautorefname{}~\ref{tab:instruction}) to initialize the Row Migration Table with source-destination row address (\texttt{src}-\texttt{dest}) pairs (\hquan{1}).
The location field of every entry is initialized to \texttt{0}.
After the Row Migration Table is populated, the CPU triggers the PIM units to perform the migration row by row (\hquan{2}).
To copy a row, the PIM unit first loads it entirely into the PIM buffer.
Upon migration completion, if the buffered data remains unmodified, the PIM unit clears the buffer and sets the corresponding location field to \texttt{1}.
Otherwise, the PIM unit re-writes the dirty buffer to the destination row.
This mechanism ensures that the CPU can always access the right data during migration.
The CPU periodically polls the Row Migration Table to monitor progress, determining that the migration is finished once all location fields are set to \texttt{1}.
Subsequently, the OS performs the necessary software updates to ensure translation consistency, which include: (1) leveraging reverse mapping (rmap) to locate all virtual pages and page-table entries (PTEs) that reference the migrated physical pages, (2) updating those PTEs to point to the new physical destination (\hquan{3}), and (3) issuing a global TLB flush as in conventional page migration (\hquan{4}).
Although the rmap phase may incur significant latency, it does not access the old physical frames, which have already been vacated, and therefore does not block the PIM program from utilizing the newly allocated chunk.
Finally, the CPU issues a \texttt{CRMT} instruction (summarized in \tableautorefname{}~\ref{tab:instruction}) to clear the entire Row Migration Table (\hquan{5}), preparing it for the next migration.

To avoid the complexity of multi-level redirection or migration chains, the system enforces serialized chunk allocation by allowing a new allocation only after the Row Migration Table is fully cleared. 
This strict ordering ensures that any subsequent memory access or migration request starts with a clean Row Migration Table state, preventing circular lookups and ensuring that the destination of one migration does not prematurely overlap with the source of another.

While this mechanism enables seamless in-DRAM redirection, a fundamental design question remains: \textit{why must this migration and redirection operate at the DRAM row level rather than the standard OS page level?}
We justify this granularity choice below.

\subsection{Justification of Migration Granularity}
\label{sec:chunk_allocation:migration_granularity}

Note that for each bank, the minimum page migration granularity of \archname{} is an entire DRAM row.
This constraint is imposed by the DRAM access protocol defined in JEDEC standards \cite{JEDEC_JESD238A_2020_ddr4,JEDEC_JESD238A_2020_lpddr5,JEDEC_JESD238A_2023_hbm3}.
During each access, the memory controller first issues an \texttt{ACT} command with the row address, which physically activates the target row.
Only afterward are column addresses sent via read or write commands to access data within the activated row.
To comply with strict DRAM timing constraints, the Row Migration Table must resolve the target physical row address (either \texttt{src} or \texttt{dest}) immediately upon receiving the \texttt{ACT} command.
Crucially, because the column address is not yet issued, the Row Migration Table cannot make decisions based on the column address. 
Otherwise, a sub-row migration would leave the Row Migration Table unable to determine the correct row to activate, inevitably causing subsequent column accesses to read corrupted data.
Thus, per-bank migration must occur at the granularity of a complete DRAM row, typically 4--16kB per rank in DDR4 (e.g., 8 devices with 0.5--2kB each) \cite{JEDEC_JESD238A_2020_ddr4}.


At the system level, per-bank migration must be coordinated across interleaved memory channels.
If only one channel's bank were migrated, a page striped across multiple channels would become only partially migrated, breaking data contiguity.
The OS must therefore coordinate migration at the granularity of a row stripe, which aggregates co-aligned physical DRAM rows across all interleaved channels (e.g., spanning 64kB in an 8-channel system with an 8kB row size).
Accordingly, \archname{} adopts this row-stripe granularity as its fundamental unit for both hardware redirection and software management.

\textbf{Feasibility of Row-Stripe-Granular Page Migration.}
Row-stripe-granularity migration implies that when allocating a PIM chunk, the OS must scan the target memory range and allocate additional row stripes to accommodate existing occupied row stripes.
Therefore, it is crucial that the system have sufficient free row stripes for allocation.
We quantify free row-stripe availability under a realistic memory fragmentation scenario.
Using a real Linux server configured as described in \sectionautorefname{}~\ref{sec:evaluation}, we executed a randomized sequence of SPEC CPU 2006 benchmarks \cite{spec200610.1145/1186736.1186737} over extended periods to induce representative memory fragmentation. 
During execution, we periodically sample the number of fully free row stripes (i.e., no page allocated within the row stripe), and plot the results in \figureautorefname{}~\ref{fig:migration}(d).
Even at high occupancy levels (e.g., >90\%), there are still over 5000 free row stripes, sufficient to accommodate the entire capacity of a chunk.
This indicates that, under typical workloads and the Linux kernel's page allocation policy, memory fragmentation smaller than a row stripe size can be effectively avoided. 
Consequently, the assumption of available destination rows for row-stripe-granular PIM-assisted migration is valid in practice.

\subsection{DRAM Instruction Support for Guaranteed PIM Chunk Allocation}
\label{sec:chunk_allocation:ins}

The instructions that support Row Migration Table management in guaranteed PIM chunk allocation are also summarized in \tablename{}~\ref{tab:instruction}.
\texttt{SRMT} appends the \texttt{src} and \texttt{dest} pairs to the Row Migration Table. 
\texttt{CRMT} is sent to clear the whole Row Migration Table after the PTEs are updated by the OS.

\section{\archname{} on Other DRAM Systems}
\label{sec:other_dram}

In this section, we generalize our dual-track memory management to other memory architectures, including HBMx and LPDDRx. 
For these architectures that lack the device-level hierarchy, we show that DRAM-side hardware can be simplified without compromising the core dual-track management.
This enables immediate deployment on existing PIM prototypes.

\begin{figure}[h]
    \centering
    \includegraphics[width=0.8\textwidth]{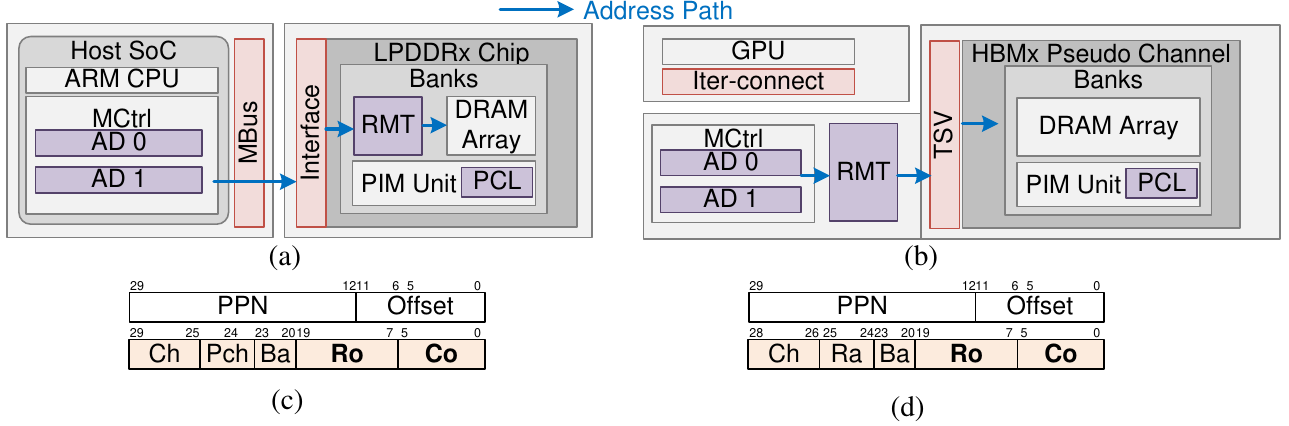}
    \vspace{-10pt}
    \caption{(a,b) Dual-track memory management on (a) LPDDRx and (b) HBMx. (c,d) Address mapping rules of PIM pages for (b) HBMx and (c) LPDDRx.}
    \vspace{-10pt}
    \label{fig:other_dram}
    \Description{Dual-track memory management on LPDDRx and HBMx}
\end{figure}

\figureautorefname{}~\ref{fig:other_dram}(a,b) depicts the architectural support for our dual-track memory management on LPDDRx and HBMx.
As both lack a device hierarchy, all remaining hierarchy levels are directly addressable by the memory controller, eliminating the need for below-bank address mapping (\sectionautorefname{}~\ref{sec:hw:mapping:intra}).
Moreover, without interleaving across devices, the normal burst layout is inherently consistent with the PIM burst, so no data re-layout hardware is required (\sectionautorefname{}~\ref{sec:hw:RC}).
In contrast, the CPU-side above-bank address mapping (\sectionautorefname{}~\ref{sec:hw:mapping:inter}) and PIM-side address translation (\sectionautorefname{}~\ref{sec:hw:pim_translate}) remain necessary.
The above-bank address mapping requires appropriate configuration of the memory controller to enable the desired address layout.
The deployment and necessity of \archname{} hardware modules are summarized in \tableautorefname{}~\ref{tab:dram-techs}.

\begin{table}[h]
\centering
\caption{Deployment and Necessity of \archname{} Components Across DRAM Technologies}
\label{tab:dram-techs}
    \vspace{-10pt}
\resizebox{0.7\linewidth}{!}{%
\begin{tabular}{|c|c|c|c|c|c|}
\hline
\multicolumn{2}{|c|}{\textbf{Technique}} & \textbf{DIMM} & \textbf{LPDDRx} & \textbf{HBMx} \\
\hline
 & Address Mapping & MemCtrl & MemCtrl &  MemCtrl \\
 &  (Above Bank) & (SoC, Config) & (SoC, Config) & (DRAM) \\ \cline{2-5}
Dual-Track Memory & Address Mapping & ATM \& RCL & / & / \\
Management &  (Below Bank) &    (DRAM)    & &  \\\cline{2-5}
 & PIM Addressing & \multicolumn{3}{c|}{PCL (DRAM)} \\\cline{2-5}
 & Data re-layout & RC (DRAM) & / & / \\\hline
\multicolumn{2}{|c|}{Guaranteed Chunk Allocation} & \multicolumn{3}{c|}{RMT (DRAM)} \\ \hline
\end{tabular}
}
    \vspace{-10pt}
\end{table}

For LPDDRx, which is commonly used in mobile SoCs, the SoC vendors typically provide configurable address decoders (AD), similar to Intel's TAD, that support multiple address-mapping regions \cite{qualcomm2026extend}.
In contrast, HBMx controllers are typically co-integrated with DRAM stacks and have fixed address mapping logic, but PIM vendors can customize them due to full control over the logic die \cite{polynesia9835628,SpaceA9407163}.
This flexibility ensures that our dual-track management can be adapted to HBMx without host-side changes.
The address mapping rules are illustrated in \figureautorefname{}~\ref{fig:other_dram}(c, d).
These rules adhere to the principle that higher-level memory hierarchies are indexed by the high-order bits of the physical address.
This mapping preserves spatial locality, ensuring that adjacent addresses map to the same memory bank whenever possible.
Thus, the dual-track memory management adapts to LPDDRx and HBMx with significantly reduced DRAM-side hardware complexity, demonstrating the generality of the design.

\section{Evaluation}
\label{sec:evaluation}

\subsection{Experiment Setup}
\label{sec:evaluation:setup}

\textbf{Baselines.} We compare \archname{} with one CPU system and two PIM system baselines on DIMM, HBM3, and LPDDR5 DRAM.
The system configurations are summarized in \tablename{}~\ref{tab:configure}.
The DRAM configuration is from the commercial DRAM PIM system UPMEM \cite{UPMEM8875680}.
In the DIMM-based PIM system configuration, the bank count is smaller than that of standard DDR4 because of the additional PIM units.
We adopt identical hardware configurations for all systems, except that half of the DRAM channel sockets are populated with PIM DRAMs in \archname{} and PIM systems.
In the \emph{CPU} system, global memory interleaving is turned on.
The \emph{PIM-Ion} system turns on global memory interleaving and uses software to translate addresses and re-layout data \cite{MetaNMP10.1145/3579371.3589091}.
The \emph{PIM-Ioff} system turns off global memory interleaving at the rank and channel levels and uses software to perform address translation and data re-layout for the remaining DRAM hierarchies, as in \cite{UPMEM8875680}.
Both PIM systems reserve a physical address space for PIM units.
\archname{} provides two address mappings for the two page types by configuring the memory controllers, as presented in \sectionautorefname{}~\ref{sec:hw:mapping:inter}.

\begin{table}[h]
    \centering
    \caption{System Configuration}
    \vspace{-5pt}
    \label{tab:configure}
\resizebox{\linewidth}{!}{%
    \begin{tabular}{cl}
        \toprule
        \multicolumn{2}{c}{Host CPU }                                           \\\midrule
        Processor   & 8-Core O3CPU @3.2GHz                                      \\
        L1/L2/L3    & 32kB, Assoc: 8 / 1MB, Assoc: 16 / 22MB, Assoc: 22. 64B Cache Line Size  \\
        \bottomrule
        \toprule
        \multicolumn{2}{c}{ DIMM DRAM Systems }                                 \\\midrule
        DRAM        & DDR4-2400, 8GB/Rank, Bank/Device/Row/Column: 8/8/131072/1024         \\
        Timing Param \cite{JEDEC_JESD238A_2020_ddr4} &
                      tBURST=3.32ns, tRCD=tCL=tRP=14.16ns, tRAS=32ns, tRRD=3.32ns,  tXAW=13.328ns, tRFC=350ns, \\
                    & tWR=15ns, tWTR=5ns, tXS=340ns, tRTP=7.5ns, tRTW=tCS=1.666ns, tREFI=7.8us                   \\
        CPU System  & 8 DRAM Channels \texttimes 4 Ranks, Addr Map: Row-Rank-Bank-Column-Channel \\
        PIM Systems & \{4 Normal, 4 PIM Channels\}
                    \texttimes 4 Ranks (PIM-Ion/off, \archname{})\\
                    & Addr Map:
                    Row-Rank-Bank-Column-Channel (PIM-Ion) / Channel-Rank-Row-Bank-Column (PIM-Off) \\\bottomrule
        \multicolumn{2}{c}{ HBM3 DRAM Systems}                                         \\\midrule
        DRAM        & HBM3-2Gbps, 1GB/Channel, Pseudo Channel/Bank/Row/Column: 2/8/32768/1024     \\
        Timing Param \cite{JEDEC_JESD238A_2023_hbm3} &
                      tBURST=4ns, tRCD=tCL=tRP=7ns, tRAS=17ns, tRRD=3ns, tFAW=15ns,tRFC=350ns, tWR=8ns, \\
                    & tWTR=4ns, tRTP=3ns, tRTW=3ns, tREFI=3.9us         \\
        PIM Systems & \{8 Normal, 8 PIM Channels\} (PIM-off, \archname{}); Addr Map: Channel-Pseudo Channel-Row-Bank-Column \\\bottomrule
        \multicolumn{2}{c}{ LPDDR5 DRAM Systems}                                        \\\midrule
        DRAM        & LPDDR5-6400, 1GB/Rank, Bank/Row/Column: 8/8/32768/1024         \\
        Timing Param \cite{JEDEC_JESD238A_2020_lpddr5} &
                      tBURST=5ns, tRCD=tRP=9.4ns, tCL=12.5ns, tRAS=21.3ns, tRRD=2.5ns, tFAW=10ns, tRFC=131ns, \\
                    & tWR=17.5ns, tWTR=6.3ns, tRTP=2.5ns, tCS=1.3ns tREFI=3.9us                   \\
        PIM Systems & \{4 Normal, 4 PIM Channels\}
                    \texttimes 4 Ranks (PIM-off, \archname{}); \\
                    & Addr Map: Channel-Rank-Row-Bank-Column (PIM-Off) \\\bottomrule
        \toprule
        \multicolumn{2}{c}{ PIM Units }                                         \\\midrule
        PIM Unit    & UPMEM DPU \cite{UPMEM8875680} @500MHz, 16 Tasklets, at Bank level \\
        \bottomrule
\end{tabular}
}
    \vspace{-10pt}
\end{table}

\textbf{Simulation.} The performance of \archname{} and baselines is evaluated with the GEM5 simulator \cite{2020arXiv200703152L} under the X86 ISA and by integrating Ramulator2 \cite{luo2023ramulator2,ramulator2_github} (based on Ramulator \cite{ramulator1_7063219}) for DRAM modeling.
We modify the address mapping in Ramulator2 and incorporate the latency of the added circuit models into the corresponding memory hierarchy, which is manifested as an increase in the read latency.
The added address mapping logic completes within two clock cycles and operates in parallel with existing decoding stages, so the total read latency increases by only two cycles.
The chunk-based non-interleaved address mapping alias of \archname{} and the reserved PIM memory spaces for PIM systems are abstracted as NUMA nodes in the GEM5 simulator.
The driver for the CPU to access the reserved memory space is ported from the UPMEM Driver \cite{UPMEM-Toolchain}.
PIM unit performance is measured with real-world UPMEM DPUs \cite{UPMEM8875680}, and the PIM task latency is inserted into the PIM-task launch point of the CPU host program to obtain the overall computing time.
The logic circuits used in the Rank Chunk List, Address Translation Module, and Re-layout Cache are synthesized through Synopsys Design Compiler with the TSMC 90nm technology library at a frequency of 2.4GHz, and the buffer array used in the Re-layout Cache is simulated with CACTI 7.0 \cite{CACTI10.1145/3085572}.
The DRAM interface is estimated with CACTI-IO \cite{cacti-io6386626}, and the DRAM die area is estimated with CACTI 7.0 \cite{CACTI10.1145/3085572}.
The software data transfer functions of baseline PIM systems used in the CPU host program, including address translation and data re-layout, are from the UPMEM Software Development Kit (SDK) \cite{UPMEM-Toolchain}.

\textbf{Benchmarks.} 
The benchmarks for evaluating \archname{} and baselines are divided into two categories.
1) \emph{CPU workloads.} CPU workloads run only on the CPU side for evaluating the impact of different memory interleaving strategies.
We select 6 benchmarks from SPEC CPU 2006 \cite{spec200610.1145/1186736.1186737}.
2) \emph{PIM workloads.} In PIM workloads, the CPU offloads memory-intensive program segments to PIM units, and other segments are executed by the host CPU.
We select 12 benchmarks from PRIM \cite{prim9771457,prim2_9651614}, GAP \cite{2015arXiv150803619B}, AiM \cite{AiMBench10.1093/bioinformatics/btad155}, and PIM-ML \cite{pim-ml10158216}.
All the benchmarks are summarized in \tablename{}~\ref{tab:benchmark}.

\begin{table}[h]

\caption{CPU and PIM Benchmarks. }
\label{tab:benchmark}
    \vspace{-5pt}

    \resizebox{\linewidth}{!}{%
    \begin{tabular}{cccc}
    \toprule
    \multicolumn{4}{c}{CPU Workloads \cite{spec200610.1145/1186736.1186737}} \\
    \multicolumn{4}{c}{(MPKI is from \cite{mpkiZENG20121029})}  \\\midrule
    Benchmark & lbm & dealII & mcf   \\
    MPKI      & 0.55 & 0.09  & 0.92  \\\midrule
    Benchmark & xalancbmk & sjeng & GCC  \\
    MPKI      &  0.13     & 0.13  & 0.03 \\
    \midrule
    \end{tabular}
    \begin{tabular}{clclcl}
    \toprule
    \multicolumn{6}{c}{PIM Workloads}  \\\midrule
    Benchmark & Parameter & Benchmark & Parameter & Benchmark & Parameter \\\midrule
    BFS \cite{prim9771457} & \#node: 196608 & UNI \cite{prim9771457} & len: 256M & HST-S  \cite{prim9771457} & len: 1.5M \\
    & sparsity: 5e-5 & TC \cite{2015arXiv150803619B} & \#node: 8k & & bins: 256 \\
    PR \cite{2015arXiv150803619B} &  Kronecker gen. & SCAN-RSS  \cite{prim9771457} & len: 8M & NW \cite{prim9771457} & size: 64k \\
    & \#node: 8k & SCAN-SSA \cite{prim9771457} & WFA \cite{AiMBench10.1093/bioinformatics/btad155} & len: 20k\texttimes{}110 \\
    MLP \cite{prim9771457} & neuron: 16k & SEL \cite{prim9771457} & len: 32M & RL \cite{pim-ml10158216} & iter: 100 \\
    & layers: 3 &&&&\\\bottomrule
    \end{tabular}
    }
    \vspace{-10pt}
\end{table}

\subsection{CPU Workloads}

\begin{figure}[h]
    \centering
    \includegraphics[width=0.55\linewidth]{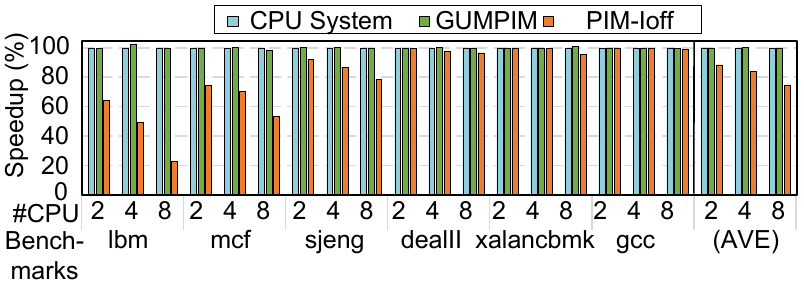}
    \vspace{-5pt}
    \caption{Performance degradation of CPU workloads on \archname{} and the PIM system compared with the CPU system.}
    \vspace{-10pt}
    \label{fig:result_cpu_task}
    \Description{}
\end{figure}

\figureautorefname{}~\ref{fig:result_cpu_task} shows the performance of CPU workloads on the CPU system, PIM-Ioff, and \archname{}, normalized to that of the CPU system.
The Cache Misses per 1000 Instructions (MPKI) of the benchmarks are summarized in \tablename{}~\ref{tab:benchmark}.
Compared with the CPU system, PIM-Ioff shows performance degradation of 88.2\%, 83.8\%, and 74.2\% for 2-, 4-, and 8-core CPU configurations across the benchmarks.
For \textit{lbm} and \textit{mcf}, which have higher MPKI, PIM-Ioff's non-interleaved address mapping exhibits >40\% performance degradation for 8-core CPUs.
When the core number increases from 2 to 8, the performance of \textit{lbm} on the PIM system decreases from 64.3\% to 22.4\%.
In contrast, for \archname{}, the performance degradation is less than 0.1\% for the benchmarks.
This is because the address interleaving of CPU pages can preserve CPU bandwidth.
The proposed hardware module for address mapping has only an ignorable impact on overall performance.
Overall, \archname{} avoids the severe CPU-side performance penalty of existing PIM systems, making it practical for multi-core environments where PIM acceleration coexists with high-bandwidth CPU workloads.

\subsection{PIM Workloads}

\begin{figure}[h]
    \centering
    \includegraphics[width=0.7\linewidth]{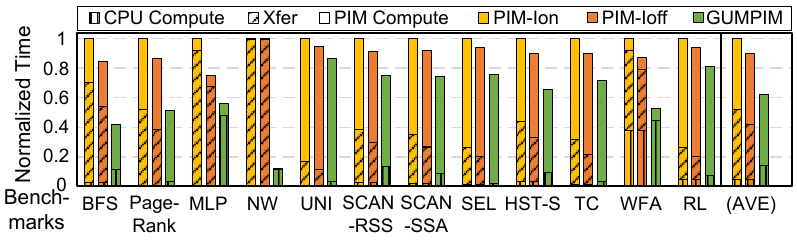}
    \vspace{-5pt}
    \caption{Computing time of PIM workloads on PIM systems and \archname{}.}
    \vspace{-10pt}
    \label{fig:result_pim_task}
    \Description{}
\end{figure}

\figureautorefname{}~\ref{fig:result_pim_task} depicts the computing time and its breakdown of PIM workloads on \archname{} and PIM systems, normalized to that of PIM-Ion.
To isolate the performance impact of dual-track memory management, we assume that the working set is preloaded in main memory, and thus exclude the overhead of memory allocation and data staging. 
This assumption aligns with standard practice in processing-in-memory research, which focuses on the efficiency of in-situ operations \cite{papi10.1145/3676641.3716009,pim-tree10.14778/3574245.3574275,pim-ml10158216,PIM-HBM9073325}.
We use 1 CPU thread, and PIM units from all 16 ranks are used.
The overall computing time is divided into three parts: CPU computing, data transfer (Xfer), and PIM unit computing.
\archname{} has no separate data transfer component because of the unitary memory space, and the CPU computing time already includes the overhead of directly accessing PIM pages.
PIM-Ioff reduces data transfer time by 29.6\% on average compared with PIM-Ion because of the simpler software-based address translation and data re-layout, demonstrating the necessity of disabling memory interleaving under isolated memory designs.
\archname{} reduces CPU computing and data transfer time by 4.93\texttimes{} on average compared with PIM-Ioff, resulting in a 1.96\texttimes{} overall workload speedup.
This reduction stems from reduced data transfer overhead and efficient address mapping and data re-layout.
Moreover, 9 of the 12 benchmarks leverage the efficient inter-PIM communication APIs, significantly reducing communication latency.
The other 3 benchmarks (BFS/PageRank/NW) cannot utilize the APIs due to their sparse, non-blocking data transfers.
However, for benchmarks with intensive, fine-grained transfers (NW/BFS/SEL in \figureautorefname{}~\ref{fig:xfer_breakdown}(a)), \archname{} still eliminates the fixed per-transfer overhead at the hardware level, boosting CPU compute speedups (8.52\texttimes{}/4.60\texttimes{}/10.65\texttimes{}).

\begin{figure}[h]
    \centering
    \includegraphics[width=0.6\linewidth]{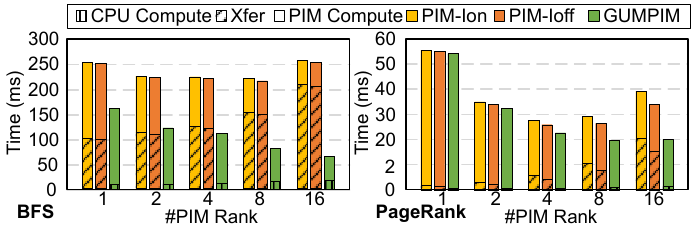}
    \vspace{-5pt}
    \caption{ Computing time of PIM workloads with different numbers of PIM units. (64 PIM units per PIM rank.)}
    \vspace{-15pt}
    \label{fig:result_pim_task_scaleout}
    \Description{}
\end{figure}

\figureautorefname{}~\ref{fig:result_pim_task_scaleout} shows the computing time of \textit{BFS} and \textit{PageRank} across different PIM unit counts.
When more PIM units are used, the amount of data transferred between PIM units increases significantly.
This results in an increase in end-to-end time when scaling beyond 8 PIM ranks, despite higher parallelism.
In contrast, the lower data transfer overhead of \archname{} makes this impact less pronounced.
As a result, \archname{} enables both benchmarks to better exploit the increased parallelism of PIM units.

\subsection{Chunk Allocation Overhead}
\label{sec:evaluation:chunk_allocation}

\begin{figure}[h]
    \centering
    \includegraphics[width=0.9\linewidth]{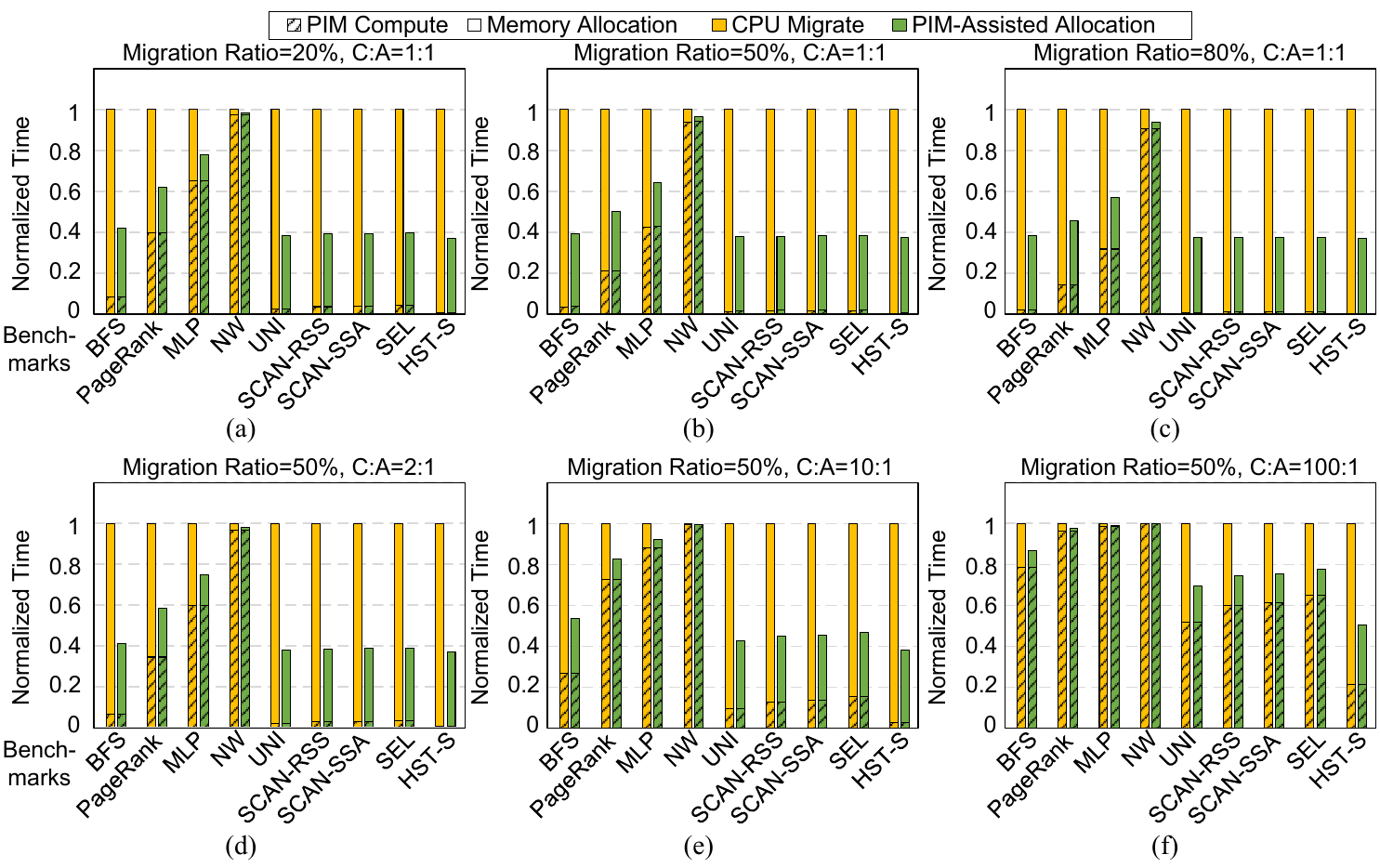}
    \vspace{-10pt}
    \caption{End-to-end time of PIM workloads with on-demand chunk allocation. ``Migration Ratio'' indicates the fraction of occupied pages within a chunk region that must be migrated to allocate it. ``C:A=2:1'' means that after the PIM pages are allocated, the PIM computation task is executed for two rounds.}
    \vspace{-10pt}
    \label{fig:result_allocation}
    \Description{}
\end{figure}

In this experiment, we explicitly include the overheads of memory allocation and data loading before the workload execution, as shown in \figureautorefname{}~\ref{fig:result_allocation}.
The page migration and memory allocation overhead is obtained using the full-system (FS) mode of GEM5, running a Linux kernel (Ubuntu 6.8.0).
We compare \archname{}'s PIM-assisted allocation mechanism against a conventional CPU-driven approach, where occupied pages are migrated by first reading them into the last-level cache and then writing them back to DRAM \cite{Contiguitas10.1145/3579371.3589079}.
\figureautorefname{}~\ref{fig:result_allocation}(a-c) compares the two approaches under different migration ratios (i.e., the fraction of occupied pages that require migration during chunk allocation) when only one round of computation follows each allocation.
The results show that the PIM-assisted allocation reduces chunk allocation time by 2.7\texttimes{}, thereby lowering the average end-to-end latency of PIM workloads by 2.3\texttimes{} at a 50\% migration ratio.
For most PIM benchmarks, including UNI, SCAN-RSS/SSA, SEL, and HST-S, chunk allocation alone accounts for over 95\% of the total end-to-end latency. 
This highlights a characteristic trait of PIM workloads, which is large memory footprints but modest computational demands. 
Consequently, for these workloads, the PIM-assisted allocation mechanism achieves an average end-to-end speedup of 2.65\texttimes{}.

We also compare the end-to-end time under different computation-to-allocation ratios (C:A) in \figureautorefname{}~\ref{fig:result_allocation}(a,d-f).
The computation-to-allocation ratio is defined as the number of computation task rounds following each allocation.
As the computation-to-allocation ratio increases, the speedup of \archname{} decreases as the proportion of computation time increases.
However, even when ``C:A=100:1'', PIM workloads with large memory footprints but modest computational demands (e.g., UNI, SCAN-RSS/SSA, SEL, and HST-S) can still benefit from the PIM-assisted allocation mechanism.
These benchmarks achieve an average speedup of 2.6\texttimes{}.
This demonstrates that reducing allocation latency remains critical for real-world PIM workloads, even when computation dominates the nominal execution profile.
The time breakdown of PIM-assisted page migration is shown in \tableautorefname{}~\ref{tab:migration_breakdown}.
The breakdown is measured by injecting M5ops \cite{gem5_m5ops}.
With PIM-assisted page migration, the data copy proportion is reduced from 58\% to less than 1\%, reducing the chunk allocation time by 2.7\texttimes{}.

\begin{figure}[h]
    \begin{minipage}{0.35\linewidth}
        \centering
        \includegraphics[width=\linewidth]{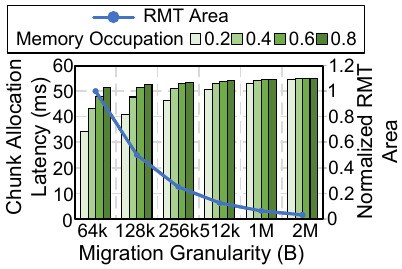}
    \vspace{-15pt}
        \caption{Row Migration Table area and chunk allocation latency under different migration granularity.}
    \vspace{-10pt}
        \label{fig:result_migration_granularity}
        \Description{result_extend}
    \end{minipage}
    \hfill
    \begin{minipage}{0.6\linewidth}
        \centering
        \includegraphics[width=\linewidth]{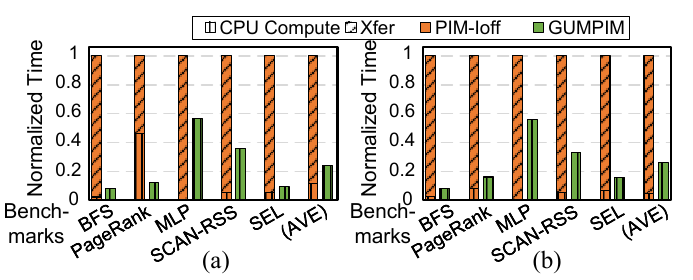}
        \vspace{-15pt}
        \caption{Speedup of \archname{} compared with PIM-Ioff on different DRAM architectures. (a) HBMx (b) LPDDRx.}
        \vspace{-10pt}
        \label{fig:result_extend}
        \Description{result_migration_granularity}
    \end{minipage}
\end{figure}

\begin{table}[h]
    \begin{minipage}{0.27\linewidth}
        \centering
        \caption{Time Breakdown of PIM-assisted Page Migration}
        \label{tab:migration_breakdown}
        \vspace{-5pt}
                
\resizebox{\linewidth}{!}{%
    \begin{tabular}{cc}
        \toprule
        Component & \begin{tabular}{c} Latency \\ (per  \\ Row Stripe) \end{tabular} \\\midrule
        Allocation & 1.7 \textmu{}s \\
        Data Copy & 0.063 \textmu{}s\\
        \begin{tabular}{c} Reverse \\ Mapping \end{tabular} & 11.6 \textmu{}s\\\bottomrule
    \end{tabular}
}
    \end{minipage}
    \hspace{0.01\linewidth}
    \begin{minipage}{0.7\linewidth}
        \centering
        \caption{Area and Power Overhead of Proposed Modules}
        \vspace{-5pt}
        \label{tab:overhead}
        \resizebox{\linewidth}{!}{%
    \begin{tabular}{|c|c|c|c|c|c|}
    \toprule
    \textbf{Position} & \textbf{Component} & \begin{tabular}{c}\textbf{Area (Overhead)} \\ \textbf{mm\textsuperscript{2}} \end{tabular} & \begin{tabular}{c} \textbf{Power(Overhead)} \\ \textbf{mW} \end{tabular} \\
    \midrule
    & PIM Chunk List & 0.032 (0.025\%) & 3.55 (1.60\%)\\
    DRAM chip & Row Migration Table & 0.134 (0.102\%) & 3.61 (1.63\%) \\\cline{2-4}
    & \textbf{Total} & \textbf{0.17 (0.13\%)} & \textbf{7.16 (3.23\%)} \\ \midrule
    & Rank Chunk List & 0.151 (0.100\%) & 3.53 (0.59\%) \\
    RCD & Address Translation Module & 3\texttimes{}10\textsuperscript{-4} (\textasciitilde{}0\%) & 0.08 (\textasciitilde{}0\%) \\
    & Command Generator + IO & 0.568 (0.376\%) & 15.95 (2.66\%) \\\cline{2-4}
    & \textbf{Total} & \textbf{0.72 (0.48\%)} & \textbf{19.48 (3.36\%)} \\\midrule
    DIMM Buffer & Re-layout Cache & 0.208 & 33.10 \\
    \bottomrule
    \end{tabular}
}

    \end{minipage}
    \vspace{-10pt}
\end{table}

\subsection{Migration Granularity}

\figureautorefname{}~\ref{fig:result_migration_granularity} plots the Row Migration Table module area and chunk allocation latency across varying migration granularities and memory utilization levels. 
Chunk allocation latency rapidly saturates at approximately 55 ms as memory utilization increases, primarily due to fragmentation effects where any occupied page necessitates migrating the entire granularity.
At high occupancy (e.g., 80\%), even finer granularities such as 64kB must migrate nearly all stripes because most row stripes contain at least one occupied page, resulting in latency similar to coarser granularities.
This saturation is exacerbated with coarser granularities; notably, at 512kB, latency reaches 50 ms even at merely 20\% memory utilization.
Conversely, finer granularities incur a significant hardware penalty, causing the Row Migration Table area to expand substantially. 
Therefore, the optimal design strategy selects the finest migration granularity that stays within the acceptable Row Migration Table area budget, effectively balancing allocation efficiency against hardware costs.

\subsection{\archname{} on LPDDR5 and HBM3}
\label{sec:evaluation:other_dram}

We also compare \archname{} with PIM-Ioff on LPDDR5 and HBM3.
The detailed configuration is summarized in \tableautorefname{}~\ref{tab:configure}.
We keep the number of banks in the LPDDR5 and HBM3 systems the same as that in the DIMM system.
As there is no device level for word expansion in LPDDR5 and HBM3, the total number of PIM units is 1/8 of that in DIMM.
The address mapping configuration is kept the same as that in the DIMM system, with higher hierarchy levels mapped to higher-order bits of the physical address.
We plot the CPU-part time, including CPU computing and data transfer, in \figureautorefname{}~\ref{fig:result_extend}.
We select 5 typical benchmarks from \tableautorefname{}~\ref{tab:benchmark}, and \archname{} reduces the CPU time by 6.0\texttimes{} on average across these benchmarks.
For PIM-Ioff on the LPDDR5 system, \archname{} reduces the CPU time by 5.9\texttimes{} on average.
The advantage is slightly smaller than that on the DIMM system because the LPDDR5 system does not have the device level, resulting in software data re-layout working at the cache-line level instead of the byte level.
Therefore, the transfer overhead is originally smaller than that in the DIMM system.
For PIM-Ioff on the HBM3 system, \archname{} reduces the CPU time by 5.2\texttimes{} on average.
This is because HBM3 DRAM originally has larger memory bandwidth, resulting in smaller transfer overhead in PIM-Ioff.
Although \archname{} provides a smaller advantage on these two DRAM systems, it can still achieve a significant speedup over PIM-Ioff.

\subsection {Area}

The area and power overhead of the proposed modules is summarized in \tablename{}~\ref{tab:overhead}.
The PIM Chunk List and Row Migration Table introduce 0.13\% overhead compared with the 130~mm\textsuperscript{2} DRAM die area derived from CACTI.
The Rank Chunk List, Address Translation Module, and Command Generator contribute 0.72~mm\textsuperscript{2} additional area (\textasciitilde{}0.48\%) to the RCD chip \cite{Renesas4RCD0232K}.
The Re-layout Cache occupies 0.21~mm\textsuperscript{2}, which is significantly smaller than the DRAM buffer chip (100~mm\textsuperscript{2}).

\section{Conclusion}
\label{sec:conclusion}

In this work, we propose \archname{}, a new system architecture with a unitary and malleable memory space for both CPU and PIM units.
\archname{} provides fine-grained interleaving for CPU pages, eliminates data transfer overhead between CPU and PIM units, and reduces PIM page allocation latency.
The method can be adapted to various DRAM architectures, including DIMM DRAM, LPDDRx, and HBMx.
\archname{} achieves only 0.1\% performance loss for CPU tasks while delivering a 2.3\texttimes{} overall speedup for PIM tasks by reducing CPU-part time and memory allocation time by 4.93\texttimes{} and 2.7\texttimes{}, respectively.

\bibliographystyle{ACM-Reference-Format}
\bibliography{refs_short3}


\end{document}